\documentclass[%
 reprint,
 amsmath,amssymb,
 aps,
pre,
showkeys,
]{revtex4-2}
\usepackage{graphicx}
\usepackage{multirow}
\usepackage{booktabs}
\usepackage{makecell}

\usepackage{subfig}
\graphicspath{ {./imgs/} }
\usepackage{bm}
\usepackage{subfig}
\usepackage{xcolor}

\newcount\myloopcounter
\newcommand{\repeatit}[2][10]{
  \myloopcounter0
  \loop\ifnum\myloopcounter < #1
  #2
  \advance\myloopcounter by 1
  \repeat
}

\begin{document}
\title{Breakdown of sequential tunnel ionization\\
in ultrashort electromagnetic pulses}

\author{D.I. Tyurin}
    \email[Correspondence email address: ]{denisturin1999@yandex.ru}
    \affiliation{National Research Nuclear University MEPhI, Kashirskoe shosse 31, 115409, Moscow, Russia}

\author{V.V. Strelkov}
    \affiliation{ P.N. Lebedev Physical Institute of the Russian Academy of Sciences, Leninsky prospect 53, 119991, Moscow, Russia}

\author{S.V. Popruzhenko}
    \affiliation{National Research Nuclear University MEPhI, Kashirskoe shosse 31, 115409, Moscow, Russia}
    \affiliation{Prokhorov General Physics Institute of the Russian Academy of Sciences, Vavilova 38, 119991, Moscow, Russia}

\date{\today} 

\begin{abstract}
 We consider double ionization of one- and two-dimensional atomic systems modeling neutral helium, xenon and the negative bromine ion in intense low-frequency electromagnetic fields.
By solving numerically the two-electron time-dependent Schr{\"o}dinger equation under conditions which exclude recollisions, we demonstrate that while for pulses of a few tens of femtoseconds duration and longer the sequential single-electron approximation perfectly describes the ionization dynamics, for pulses as short as few femtoseconds this picture breaks down entirely, and the electron-electron interaction suppresses the rate of ionization by roughly one order of magnitude.
\end{abstract}


\maketitle

\section{Introduction}

The fundamental effect of quantum mechanical tunneling discussed for the first time in the context of $\alpha$-decays \cite{gamow-28,condon-pr29} and ionization of hydrogen in a weak static field \cite{opp_pr28} is pivotal in the physics of nuclei \cite{weisskopf-book}, the physics of atoms in external fields \cite{fedorov-book,popov-usp04}, and the physics of condensed matter \cite{ankerhold-07}. 
Optical tunneling in atoms, molecules, and solids  underlies attosecond science \cite{ivanov-rmp09,calegari-adv16}, the generation of high-order harmonics (HHG) \cite{gaarde-josab22,strelkov-usp23}, and attosecond spectroscopy \cite{agostini_nobel24,huillier_nobel24,krausz_nobel24}.
Tunnel and multiphoton ionization of atoms in the field of intense laser radiation first theoretically addressed by Keldysh \cite{keldysh-jetp65}, was observed initially as multiquantum absorption \cite{delone-jetp66}, and later -- in the form of above-threshold ionization (ATI) \cite{agostini-prl79}.
The effects of ATI and HHG can be efficiently explained within the semi-classical three-step model \cite{kuchiev-jetpl87, corkum-prl93}, where optical tunneling serves as the first step.

A wide range of theoretical approaches describing the phenomenon of strong-field ionization and related processes are based on the single-active electron (SAE) approximation.
This is commonly substantiated by the fact that, in the regime of optical tunneling, the electron must travel a sufficient distance from the nucleus to appear in that part of space where the laser force dominates over the Coulomb attraction.
The higher the ionization potential $I_p$ of the level, the smaller is the probability that the electron appears at large distances from the nucleus, so that the model in which the currently outermost electron solely interacts with the laser field sounds adequate.
The celebrated exclusion from this rule takes place in linearly polarized fields of non-relativistic intensity, where such an electron may undergo a recollision with the atomic residual along the same scenario, as underlies the effects of HHG \cite{corkum-prl93} and high-energy ATI plateau \cite{paulus-prl94}.
Recollision leads to an extreme enhancement in double ionization (DI) that is up to several orders of magnitude, compared to that expected for the sequential mechanism.
This effect of nonsequential double ionization (NSDI) is the eminent example of an  ionization channel strongly mediated by the electron-electron interaction in the regime of tunneling.
By now, NSDI in atoms has been thoroughly studied both experimentally and in theory so that its recollision genesis is beyond doubt; see Ref.\cite{becker-rmp12} for a review.
Other nosequential ionization mechanisms including the shake-off model \cite{agostini-rev95} and the collective tunneling model \cite{zon-jetp99,becker-prl00,tyurin-leb23} have been theoretically and numerically examined \cite{lein_prl00, becker-prl00} but not yet observed in experiments.

In linearly polarized fields of non-relativistic intensity, the NSDI through recollision dominates other possible DI channels by orders of magnitude, which makes a search for alternative mechanisms extremely difficult both in numerical calculations and in experiment.
Moreover, the NSDI associated with recollisions appears a remarkably broad concept, which allows also entirely classical interpretations \cite{eberly_prl05,mauger_prl12,huang_oe13} and covers, in a row with the plane hard recollision followed by the immediate release of the second electron \cite{kopold_prl00}, several indirect sub-channels.
In particular, the second electron can be emitted through the excitation of a bound state followed by field ionization \cite{rudenko_prl04}, through soft recollisions \cite{emma_pra09}, or the excitation into continuum can be facilitated by the complex dynamics of the laser-dressed outer electron \cite{mauger_prl12}.
There are three DI regimes where  alternatives of the recollision scenario can be revealed: (i) relativistic ionization, (ii) ionization in fields with circular or elliptical polarization where rescattering is damped out due to the two-dimensional motion of the laser-driven electron (see e.g. \cite{keitel-pra07,kolya-pra08} for details and examples), and (iii)  ionization by extremely short laser pulses with duration insufficient to bring the electron back to its parent ion.
NSDI in ultra-short laser pulses has experimentally demonstrated a transition between regimes with few and multiple recollisions \cite{kubel_njp14}.  
Even shorter pulses supporting {\em no recollision} are hardly achievable in experiments, but they are easy to reach by modeling strong-field ionization in unipolar fields.
This allows for addressing tiny effects otherwise shielded by the oscillating electron's dynamics \cite{keitel-pra22}.

For the following, some elaboration of the terminology is required to avoid ambiguities and misinterpretations.
We characterize ionization as sequential if the electrons escape from the atom one after another with some significant time interval. 
The electron-electron interaction may or may not play an important role in this process, so that the sequential ionization is not necessarily equivalent to ionization of non-interacting electrons.
The nonsequential ionization is by definition something opposite, so it is a process in which both electrons depart from the ion within a small time window. 
Note that in the recollision induced NSDI the first electron leaves the atom approximately one laser period earlier than the second one. However, this fact does not withdraw this channel from the present definition, because eventually the electrons depart simultaneously, after the recollision, while the preceding evolution of the first electron is a virtual process. 
The term NSDI is equally applicable to different mechanisms of DI \cite{kulander_prl92,moshammer_prl01,becker-prl00,lein_prl00}.  
Simultaneous emission can take place for non-interacting electrons or for interacting ones, but without recollision and excitation of intermediate bound states.

In this paper, we examine the breakdown of the sequential single-active-electron picture of double ionization of atoms and negative ions in strong ultrashort laser pulses under conditions {\em entirely eliminating recollision}. 
The Keldysh parameter (here and below we use atomic units $m=e=\hbar=1$)
\begin{equation}
    \gamma=\frac{\sqrt{2I_p}\omega}{E_0}
    \label{gamma}
\end{equation} 
\cite{keldysh-jetp65} is kept about unity or smaller, so that the ionization proceeds through tunneling rather than through multiphoton absorption. 
Here $\omega$ and $E_0$ are the characteristic frequency and strength of the electric field, $I_p$ is the ionization potential.

We consider three targets: helium, the simplest two-electron atom in its singlet state; xenon, the atom extensively used in experiments on strong-field DI, and the negative bromine ion, the target specific due to the great difference in the first and second ionization potentials. The latter makes negative ions promising candidates for the search of collective ionization, which can be one of channels alternative to sequential ionization \cite{becker-prl00}. 
For all three systems, we solve the time-dependent Schr\"odinger equation (TDSE) for one- and two-dimensional models (see Sec.~\ref{sec:method} for details) and calculate the double ionization probability. 
 For He, the full two-dimensional calculation exceeds our computational resources, so only 1D results are presented. 
We show that, in short pulses of a few-femtosecond duration, the double ionization yield deviates dramatically from that expected in the sequential picture, while in longer pulses of several tens of femtoseconds the SAE approximation reproduces the exact numerical result with remarkable accuracy. 
We demonstrate that this breakdown is caused by the electron-electron repulsion, which suppresses the sequential ionization channel by about an order of magnitude in the short-pulse regime. This suppression is present both in 1D and 2D systems and for all three targets, indicating that the general nature of the effect.

The paper is organized as follows.
In Section~\ref{sec:method}, we introduce basic equations and explain our numerical scheme.
Section~\ref{sec:results} is dedicated to the analysis of numerical results: first we define the long- and short-pulse regimes and introduce a quantitative separation criterion, then we demonstrate the validity of the SAE approximation in the long-pulse regime and its breakdown in the short-pulse regime in 1D and 2D geometry.
The last section contains conclusions.
Appendix~\ref{ap:A} introduces analytic single-electron ionization rates used for the analysis of the sequential ionization dynamics in terms of rate equations.

\section{Statement of the problem, basic equations and numerical method}\label{sec:method}

 To investigate the influence of the electron-electron interaction on the double ionization yield, we compare the ionization dynamics of two model systems: (a) a two-electron atom or a negative ion  with interacting electrons and (b) an atom or ion with two non-interacting electrons having the same ionization potentials as (a).
In case (a), interaction with the parent ion $V_\text{ei}$ and electron-electron interaction $V_\text{ee}$ are described by the smoothed Coulomb potentials  widely used for modeling of atomic systems driven by intense electromagnetic fields \cite{hall2009energies, fernandez1991analytical,lima1981atoms, mehta1978bound, silaev2010strong, gebremedhin2014calculations}
\begin{equation}\label{Vion}
    V_\text{ei}(\bm r) = \frac{-Z}{\sqrt{\bm r^2 + a^2}}~,
\end{equation}
\begin{equation}\label{Vint}
    V_\text{ee}(\bm r_1, \bm r_2) = \frac{1}{\sqrt{(\bm r_1 - \bm r_2)^2 + b^2}}~,
\end{equation}
 where $Z$ is the residual atomic charge after the removal of the two external electrons. 
For a negative ion $Z=1$, for the atom $Z=2$.
The smoothing parameter $a$ is chosen to reproduce the second ionization potential $I_{p2}$ when the outermost electron is removed.  
After this, the parameter $b$ is adjusted to reproduce the $I_{p1}$ of the outer electron. 
Parameters $a$ and $b$ and the ground-state wavefunction are found by solving numerically the TDSE.
 Note, that parameters $a$ and $b$ are different for 1D and 2D cases.

The two-electron TDSE for the case (a) reads 
\begin{equation}\label{2eTDSE}
\begin{gathered}
    i\frac{\partial}{\partial t} \Psi(\bm r_1, \bm r_2, t) = \\
    = \big[H_0(\bm r_1, t) + H_0(\bm r_2, t) + V_\text{ee}(\bm r_1, \bm r_2)\big]\Psi(\bm r_1, \bm r_2, t)~,
\end{gathered}
\end{equation}
where $H_0$ is the single particle Hamiltonian including the interaction with the electric field of the pulse
\begin{equation}\label{H0}
    H_{0}(\bm r, t) = -\frac{1}{2} \Delta + V_\text{ion}(\bm r) - \varphi (\bm r, t)~
\end{equation}
with $\varphi$ being its scalar potential. 
To suppress recollisions, the time dependence of the electric field $E(t)$ is set in the form of an unipolar pulse with the amplitude value $E_0$ and the full duration $T$ 
\begin{equation}\label{E(t)}
    E(t) = -E_0\sin^2\left(\dfrac{\pi t}{T}\right)~.
\end{equation}
In Eqs. (\ref{2eTDSE}), (\ref{H0}), the interaction of the electrons with the field of the unipolar pulse is introduced through the scalar potential: $\varphi (\bm r, t) = -\int \tilde E (x, t)dx$, here $\tilde E (x, t)$ is the electric field linearly polarized in the x-direction
In a spatially homogeneous field as (\ref{E(t)}), the electrons are accelerated after ionization, rapidly reaching the boundaries of the computation box.  
This hinders analysis of the two-electron ionization dynamics. 
To make such analysis more affordable, in the numerical calculation we use the field which is close to homogeneous in the vicinity of the ion and vanishes far from it:
\begin{equation}\label{E(r,t)}
\tilde E(x, t) = 
   \begin{cases}
       E(t)\cos^2\left(\dfrac{\pi x}{2 x_0}\right),~|x|\leq x_0 \\
       0,~\text{elsewhere}~,
   \end{cases}
\end{equation}
where $x_0=30$ a.u.
With this choice, the field remains almost spatially homogeneous on the scale of the tunnel barrier $b=I_p/E_0\simeq 10$ a.u.
 For the 1D case, the sensitivity of results to the value of $x_0$ has been checked by increasing it up to $x_0=300$a.u., which practically corresponds to a homogeneous field. 
The results become insensitive to the value of $x_0$ for $x_0>100$a.u., while for smaller values the total ionization probability appears suppressed by some factor, which is of minor significance compared to the effect we discuss below.

The TDSE (\ref{2eTDSE}) is solved, using the split-step Fourier method \cite{weideman1986split}.
We consider Xe and $\text{Br}^-$ both in 1D and 2D, while for He, the 2D two-electron calculations lie beyond our computational resources, so that we restrict the He simulations to two 1D electrons. 
The electric field amplitude $E_0$ in Eq.(\ref{E(t)}) is chosen as $0.35$ for He, $0.06$ for Xe, and $0.034$ for Br$^-$. 
These values ensure that the Keldysh parameter (\ref{gamma}) remains in the tunneling or intermediate regime ($\gamma\le 1$) for all considered pulse durations.

For the 1D case, coordinates $x_1$ and $x_2$ of the electrons were discretized on a spatial grid of $2048\times 2048$ nodes. 
The spatial and time discretization steps $\Delta x$ and $\Delta t$ are presented in Table~1.  
We have performed numerical checks showing that the reduction of $\Delta x$ or $\Delta t$ below these values does not affect the results.
For the 2D case, the coordinates were discretized with a step $\Delta x = \Delta y$ same as in 1D (see Table.\ref{tab:discretization}) using $N_x = 512$, $N_y = 256$ nodes for each electron.
The total size of the computational box is $512\times256\times512\times256$ nodes.

\begin{table}[h]
    \centering
    \caption{Time and space discretization steps}
    \label{tab:discretization}
    \begin{tabular}{|l|c|c|}
        \hline
         & $\Delta t$ (a.u.) & $\Delta x$ (a.u.)\\
        \hline
        He    & 0.02 & 0.2 \\
        Xe    & 0.1  & 0.3 \\
        Br$^-$ & 0.2  & 0.5 \\
        \hline
    \end{tabular}
\end{table}

The absorbing layer \cite{neuhasuer1989} occupies 15 nodes at the boundary of the computational box. Note that the split-step Fourier method requires a rectangular computational grid. Therefore, the double ionization probability cannot be obtained as the part of the wave function absorbed at the boundaries, as is done, for example, in \cite{scrinzi2012}. 
For this reason, we choose the computational box large enough so that during the propagation the wave function does not reach the boundaries. Adding an absorbing layer at the boundary is only necessary to control this fact by monitoring the conservation of the wave function norm.

To reproduce the first $I_{p1}$ and second $I_{p2}$ ionization potentials of our model atomic systems we used the smoothing parameters $a$, $b$ in (\ref{Vion}) and (\ref{Vint}), presented in Table.\ref{tab:interacting_parameters}.

\renewcommand{\arraystretch}{1.2}
\begin{table}[h]
\centering
\caption{Parameters of the potentials (\ref{Vion}) and (\ref{Vint}) for the model with two interacting electrons.}
\label{tab:interacting_parameters}
\begin{tabular}{|l|c|c|c|c|c|c|c|}
\hline
\multirow{2}{*}{} & \multirow{2}{*}{$I_{p1}$ (eV)} & \multirow{2}{*}{$I_{p2}$ (eV)} & \multirow{2}{*}{$~~~Z~~~$} & \multicolumn{2}{c|}{1D} & \multicolumn{2}{c|}{2D} \\
\cline{5-8}
 & & & & $a$ & $b$ & $a$ & $b$ \\
\noalign{\hrule height 1.5pt}
Br$^-$ & 3.37 & 11.81 & 1 & 1.660 & 2.600 & 1.000 & 2.200 \\
\noalign{\hrule height 1.5pt}
Xe & 12.13 & 21.21 & 2 & 2.065 & 2.520 & 1.546 & 2.100 \\
\noalign{\hrule height 1.5pt}
He & 24.59 & 54.42 & 2 & 0.707 & 0.577 & --- & --- \\
\hline
\end{tabular}
\end{table}

In the case (b) of the system with non-interacting electrons, the TDSE with the single-particle Hamiltonian was solved for the first and second electrons separately. 
For the outer and inner electron in He and Xe, we used the smoothed Coulomb potential (\ref{Vion}) with $Z=1$ and $Z=2$ for the outer and inner electron, correspondingly. 
For the outer electron in the negative bromine ion, the potential was approximated by a short-range well of the form
\begin{equation}\label{Vshort}
    V_\text{short}(r) = -c_1\exp\left(-\dfrac{r^2}{c_2^2}\right)
\end{equation}
with two fitting parameters $c_1$ and $c_2$ chosen to reproduce the first ionization potential $I_{p1}$. 
For the second electron in $\text{Br}^-$, we used the smoothed Coulomb potential (\ref{Vion}) with $Z=1$ and a fitting parameter $a$ chosen to reproduce $I_{p2}$.
The fitting parameters are presented in Table.\ref{tab:noninteracting_parameters}.

\renewcommand{\arraystretch}{2}
\begin{table}[h]
\centering
\caption{Parameters of the single-electron potentials used in the model with non-interacting electrons.}
\label{tab:noninteracting_parameters}
\begin{tabular}{|l|l|l|c|c|}
\hline
\multirow{2}{*}{} & \multirow{2}{*}{\makecell{Ionization \\potentials [eV]}} & \multirow{2}{*}{\makecell{Bound\\potential}} & \multicolumn{2}{c|}{Parameters} \\
\cline{4-5}
 & & & 1D & 2D \\
\noalign{\hrule height 1.5pt}
\multirow{2}{*}{Br$^-$} & \makecell{$I_{p1}=3.37$} & Eq.~(\ref{Vshort}) & \makecell{$c_1=0.48$\\$c_2=0.76$} & \makecell{$c_1=0.48$\\$c_2=2.0$} \\
\cline{2-5}
 & \makecell{$I_{p2}=11.81$} & Eq.~(\ref{Vion}) & \makecell{$Z=1$\\$a=1.66$} & \makecell{$Z=1$\\ $a=1.0$} \\
\noalign{\hrule height 1.5pt}
\multirow{2}{*}{Xe} & \makecell{$I_{p1}=12.13$} & Eq.~(\ref{Vion}) & \makecell{$Z=1$\\ $a=1.617$} & \makecell{$Z=1$\\ $a=0.961$} \\
\cline{2-5}
 & \makecell{$I_{p2}=21.21$} & Eq.~(\ref{Vion}) & \makecell{$Z=2$\\ $a=2.065$} & \makecell{$Z=2$\\ $a=1.546$} \\
\noalign{\hrule height 1.5pt}
\multirow{2}{*}{He} & \makecell{$I_{p1}=24.59$} & Eq.~(\ref{Vion}) & \makecell{$Z=1$\\ $a=0.695$} & --- \\
\cline{2-5}
 & \makecell{$I_{p2}=54.42$} & Eq.~(\ref{Vion}) & \makecell{$Z=2$\\ $a=0.707$} & --- \\
\hline
\end{tabular}
\end{table}

In both cases (a) and (b) the initial wave function $\Psi_0(\bm r_1, \bm r_2, t)$ was symmetrized with respect to the coordinates assuming zero total spin.
 For helium, this corresponds to the singlet ground state, while for the two other systems this should be considered as a model anzatz, which leaves the role of angular momentum and spin of the initial two-electron state beyond our consideration. 
The two-electron ground state wave function was obtained using the imaginary time method \cite{bader2013}.

The double ionization probability $W(t)$ was found by integrating the modulus squared wave function over space with the ion vicinity excluded
\begin{equation}\label{W(t)}
    W(t) = \int \limits_{
    \substack{
    |\bm r_{1}|>\varkappa\\
    |\bm r_{2}|>\varkappa
    }
    } |\Psi(\bm r_1, \bm r_2, t)|^2 d\bm r_1 d\bm r_2~.
\end{equation}
Here, $\varkappa=20$ is the size of the area around the ion, which was selected such that the probability $W$ does not depend on $t$ at the end of the calculation.

To represent the space probability distribution for 2D systems on the plane $(x_1, x_2)$ of the coordinates along the field direction, we integrate $|\Psi({\bf r}_1,{\bf r}_2,t)|^2$  over the transverse coordinates $y_1$ and $y_2$ and plot the function:
\begin{equation}\label{Fx1x2}
    F(x_1, x_2, t) = \int
    |\Psi(x_1, y_1, x_2, y_2, t)|^2 dy_1 dy_2~.
\end{equation}

\section{Double ionization in long and short pulses}\label{sec:results}

\subsection{Long- and short-pulse regimes: a quantitative criterion}

We consider two ionization regimes, qualitatively distinguished by the role of electron-electron correlations in double ionization.
In the first case, which we refer to as the long-pulse regime (LPR), the pulse duration is sufficiently long for the ionization probability of the outer electron to saturate on the front of the pulse, before the field becomes strong enough to affect the inner electron.
Thus, by the time the ionization probability of the inner electron becomes significant, the outer orbital is already depleted.
In this case, the electron-electron interaction is significant only in the initial bound state, and the subsequent ionization dynamics is expected to be well described by the SAE approximation.
In the opposite short-pulse regime (SPR), the pulse duration is shorter than the saturation time of the outer electron.
As a consequence, by the time the field becomes strong enough to ionize the inner orbital, the outer electron has not yet left the vicinity of the ion.
In this case, the electron-electron interaction plays an important role when one or both electrons entered continuum, and therefore it can significantly modify the double ionization yield.

To establish a quantitative criterion separating the two regimes, we model the sequential ionization dynamics using the time-dependent quasistatic ionization rates known as Smirnov--Chibisov (SC) or Perelomov--Popov--Terent'ev (PPT) formulas \cite{smirnov_jetp66,popov_jetp66,perelomov_jetp67} (see also reviews \cite{popov-usp04,poprz_jpb14}).
Details of these calculations are given in Appendix~\ref{ap:A}.
The border between the regimes is determined by the ionization saturation time $t_s$ of the outer electron.
Quantitatively, we define $t_s$ as the pulse duration for which the population $n_1$ of singly ionized atoms reaches $0.8$ at the field maximum.
If $T>t_s$, the outer orbital is almost entirely depleted before the inner electron is significantly affected; if $T<t_s$, the inner electron begins to feel the field while the outer electron is still bound.
Numerical analysis based on the rate equations gives $t_s\approx 12$~fs for both Br$^-$ and Xe, while for He the saturation time is considerably shorter, $t_s\approx 3$~fs, owing to its higher ionization potentials.
Thus, for Xe and Br$^-$ one could expect realization of the SPR in pulses of few-fs duration accessible for the modern high-power optical or infrared laser systems.
For He, a similar regime would require even shorter pulses.

Note that the value of $t_s$ defined as described above (see also Appendix A) gives only a rough estimate based on the rate-equation model and therefore calls for further substantiation. 
In the following subsection, we turn to the solution of the two-electron TDSE and present \textit{ab initio} results which demonstrate suppression of double ionization in the short-pulse regime, in agreement with the qualitative picture sketched above.

\subsection{Suppression of double ionization in 1D and 2D models}

\begin{figure}[h]
    \centering
    \includegraphics[width=0.85\linewidth]{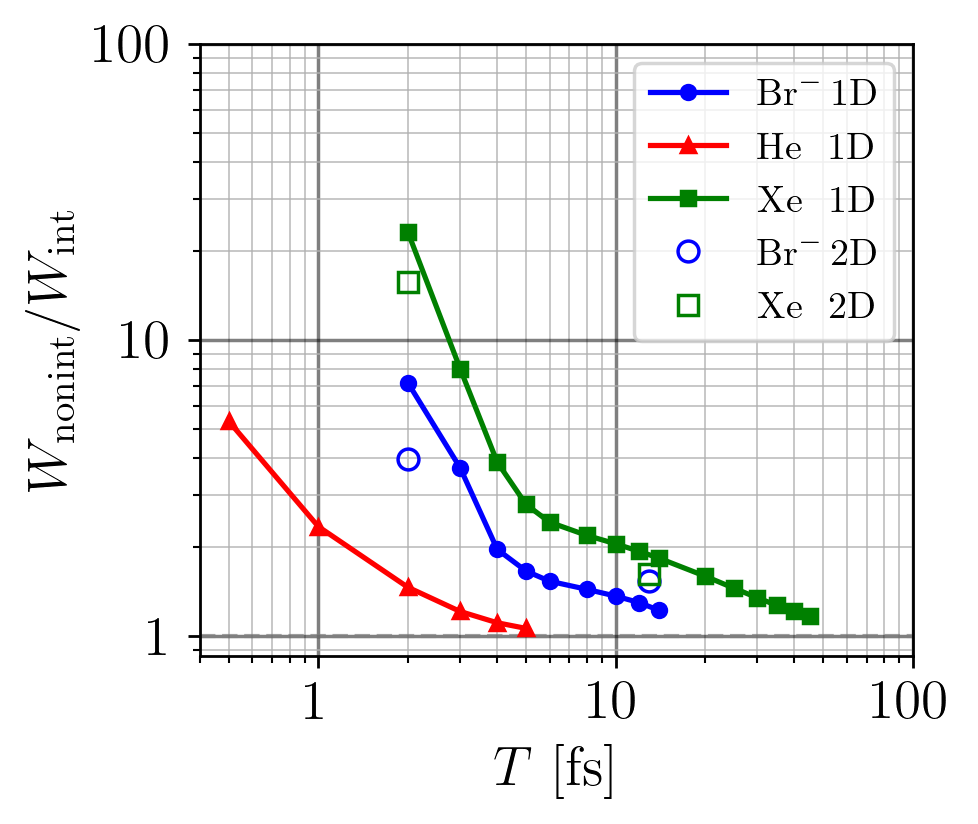}
    \caption{
     Ratio of the DI probabilities obtained without and with the electron-electron interaction, $W_{\rm{nonint}}/W_{\rm{int}}$, as a function of the pulse duration $T$ for He, Xe, and Br$^-$. Solid lines show the 1D results. Open squares and open circles mark the 2D results for Xe and Br$^-$, respectively, at the two pulse durations $T=13$ fs and $T=2$ fs. Note log vertical and horizontal scales.}
    
    \label{fig:W_ratio}
\end{figure}

To perform a scan over the pulse duration $T$ for all three targets and to verify that $t_s$ indeed separates the two regimes, we begin with computationally cheap 1D calculations.
For each value of $T$, we solve the two-electron TDSE with and without the electron-electron interaction. 
The resulting DI probabilities, calculated from Eq.~(\ref{W(t)}), will be referred to as $W_{\mathrm{int}}$ and $W_{\mathrm{nonint}}$. 
We compare these probabilities once they have saturated and no longer depend on time.
The results are presented in Fig.\ref{fig:W_ratio}.
For $T>t_s$ the ratio $W_{\mathrm{nonint}}/W_{\mathrm{int}}$ remains close to unity, confirming that the interaction does not affect the double ionization yield: this is precisely the regime in which the SAE approximation is justified.
As $T$ decreases below $t_s$, the ratio grows rapidly, demonstrating a strong suppression of sequential ionization. 
For the shortest pulses the ratio is close to one order of magnitude for all three targets, indicating the general nature of this effect.

In the 1D model, however, the electrons are strictly confined by collinear geometry, which can lead to an overestimation in the suppression effect.
In order to demonstrate that the effect is weakly sensitive to the dimensionality, we turn to 2D systems.
Here, the available computational resources do not allow us to perform calculations for the same wide range of pulse durations as in 1D.
We therefore choose two representative pulse durations: $T=2$~fs, lying well within the SPR, and $T=13$~fs, the largest duration we can afford numerically. 
The latter is comparable to the saturation time $t_s\approx 12$~fs and therefore serves only as a reasonable approximation to the LPR.
For He, whose significantly larger ionization potentials would require a much finer spatial grid and considerably shorter time steps, fully two-dimensional calculations lie beyond our current computational resources. 
We therefore demonstrate the 2D effect for Xe and Br$^-$.

\begin{figure}[h]
    \centering
    \includegraphics[width=1\linewidth]{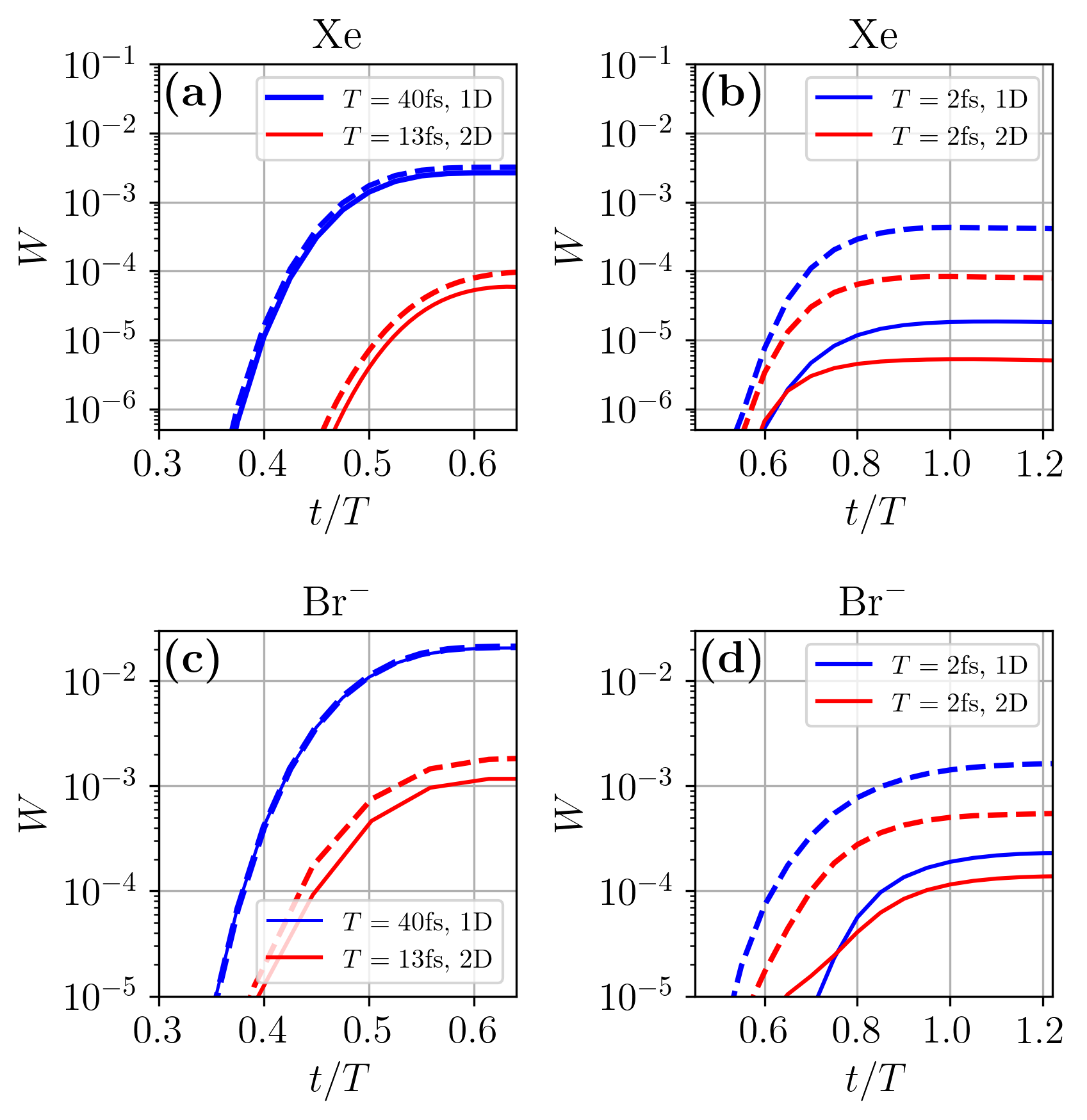}
    \caption{
    Total probability of double ionization (\ref{W(t)}) for Xe (a, b) and $\text{Br}^-$ (c, d).
    Blue and red lines correspond to the 1D and  2D cases respectively. Solid lines show the case of the interacting electrons, dashed lines -- that of the non-interacting electrons. 
    Panels (a,c): LPR, with $T=13$~fs for 2D and $T=40$~fs for 1D.
    Panels (b,d): SPR, with $T=2$~fs for both 2D and 1D.
    }
    \label{fig:2D_W}
\end{figure}

Fig.\ref{fig:2D_W} shows the total probability of double ionization $W$ as a function of time for Xe and Br$^-$ in 2D.
For $T=13$ fs (panels (a, c)), the probabilities found with and without the electron-electron interaction are rather close to each other.
It is natural to expect that as the pulse duration increases, the accuracy of the SAE approximation without the electron-electron interaction included will only improve.
To demonstrate this, we present on the same plots the 1D calculations for $T=40$~fs (panels (a,c), blue lines), where the interacting and non-interacting results are nearly indistinguishable.
In contrast, for $T=2$ fs (panels (b,d)) the interaction reduces the double ionization yield by approximately one order of magnitude for Xe and Br$^-$, in both the 1D and the 2D cases.
This demonstrates that the suppression of sequential ionization is not specifically inherent to 1D geometry.

\begin{figure}[h]
    \centering
    \includegraphics[width=1\linewidth]{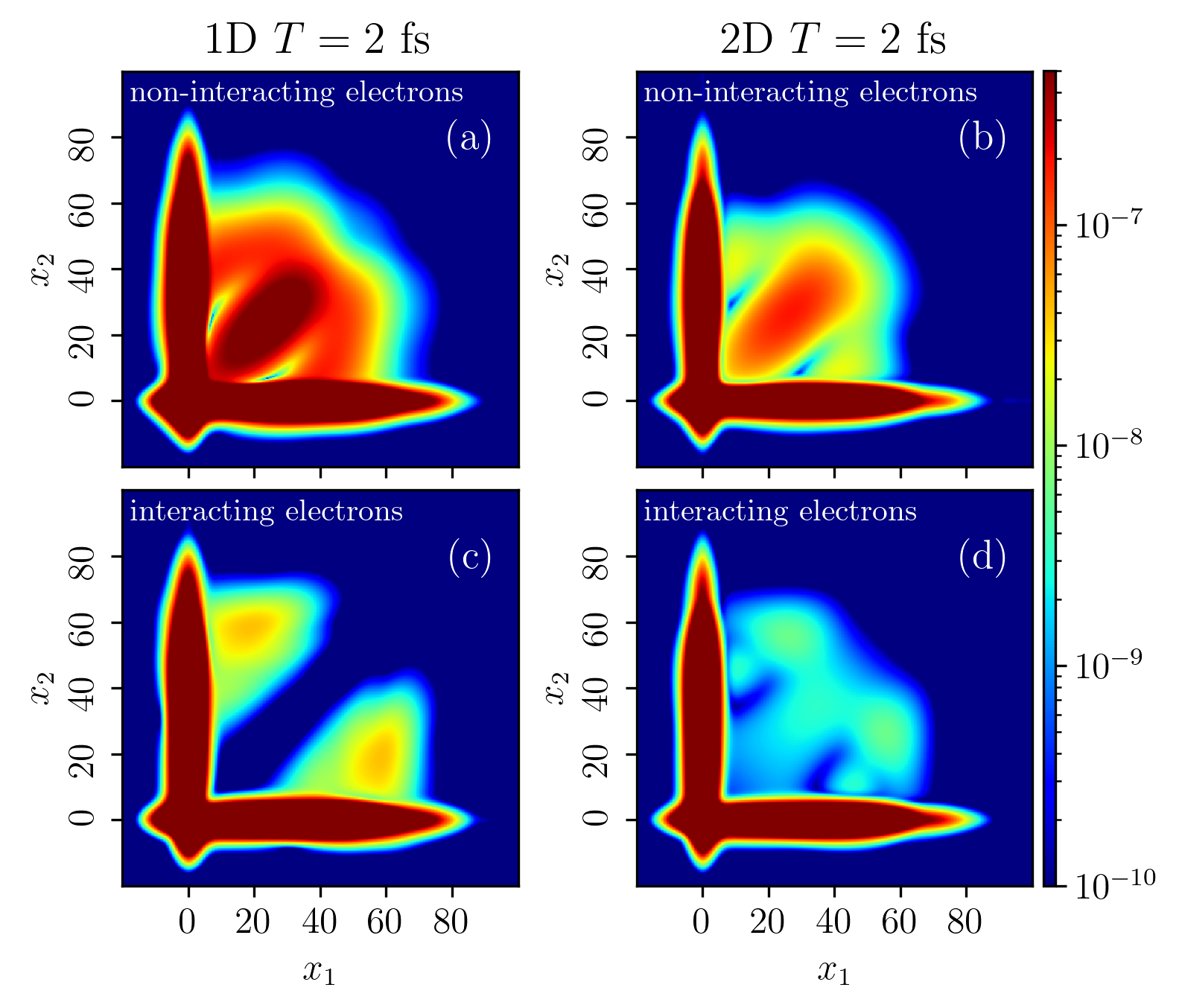}
    \caption{
     Two-electron probability distributions for Xe in the SPR ($T=2$~fs). Upper row: 1D~(a) and 2D~(b) distributions for the model with the noninteracting electrons. Lower row: 1D~(c) and 2D~(d) distributions for the model with the interacting electrons. The distributions are
    shown at the end of the pulse ($t=T/2$).}
    
    \label{fig:Xe_SPR_density}
\end{figure}

To further examine the effect, in Fig.\ref{fig:Xe_SPR_density} we compare the two-electron probability distributions for the interacting and non-interacting electrons in 1D and 2D Xe considering the SPR ($T=2$~fs).
For the non-interacting electrons (panels (a,b)), a significant part of the density is concentrated near the diagonal $x_1=x_2$, indicating that the two electrons tend to escape simultaneously.
When the electron-electron interaction is taken into account (panels (c,d)), this diagonal flux is highly suppressed.
In the 1D case, the suppression is almost absolute, while in the 2D model a visible diagonal flux survives, although it is significantly reduced compared with the non-interacting case.

The total suppression of the diagonal flux in 1D can be qualitatively explained by the following naive picture.
Let two bound electrons be shifted by the electric field along its direction, and their positions are different.
The electron, which is closer to the core, is pulled back not only by the nucleus attraction but also by the repulsion of the other electron.
This repulsion reduces the probability of detachment for the inner electron until the outer one is far from the atom.

In 2D, the diagonal flux survives.
The reason for its persistence is that in 2D an additional collective ionization channel becomes available, which is forbidden in 1D.
A detailed study of this channel is presented in Ref.\cite{tyurin_arxiv26}.

\section{Conclusions}

 In conclusion, we numerically examined double ionization of 1D and 2D two-electron systems, which model the negative bromine ion, neutral xenon and neutral helium atoms.
Our study has revealed conditions at which the e-e interaction influences the DI rate, destroying the generally accepted picture of sequential single-electron tunneling.
This happens in short pulses, when the ionization probability for the inner electron becomes sizable before the outer electron is removed with guarantee.
In this case, the DI probability appears suppressed by about one order of magnitude, compared to the case of independent sequential tunneling.
Thus, the single active electron approximation customary in the theory of strong field processes, is highly limited for intense ultrashort laser pulses.
 For a two-electron system with a symmetric ground state as we consider in our calculation, helium in its singlet state is formally the most relevant object.
However, with the ionization potential growing, the border between the regimes of long and short pulses shifts toward shorter pulse durations, and a sizable breakdown of the sequential ionization picture happens there for pulses shorter than a femtosecond.
Such ultra-short pulses are not available in the optical and infrared spectral domains, where the tunnel regime of ionization can be achieved.
This makes He the system less interesting for an experimental search of the predicted effect than atoms with smaller ionization potentials, viz. xenon.
In negative ions, where the first and the second ionization potentials differ considerably, the breakdown appears comparably strong as in Xe.
We also note that in the 2D case with interacting electrons, a diagonal two-electron flux emerges, indicating a collective ionization channel in which two electrons escape from the atom side by side.
We will consider this channel of correlated nonsequential ionization which can not be attributed to the recollision mechanism in a separate study \cite{tyurin_arxiv26}.

 A phenomenon close in the underlying physics to that we study in this manuscript has been examined theoretically by Bart with co-authors \cite{bart_photonics20} for sub-nm clusters subject to strong single-cycle near-infrared pulses.
A significant difference between the atomic systems we address here and small clusters studied in \cite{bart_photonics20} is that, in the latter case, the single-electron levels are rather dense, and the ionization potentials appear close to each other. 
In clusters, the effect was called the Coulomb blockade, and the suppression of ST by 2-3 orders of magnitude was demonstrated. 
Thus, in our calculations we show that a similar mechanism can generate a strong (although smaller than in clusters) effect in atoms and negative ions with highly different ionization potentials of the first and second electrons.

The predicted breakdown of the sequential tunneling picture should play a quantitatively important role in the process of multiple ionization of atoms in short laser pulses of extreme intensities, where, due to the high peak field and the steepness of the time envelope the ionization saturation time could lie in the sub-fs domain.
In particular, the suppression of ionization induced by the e-e correlation should be pivotal for the development of numerical particle-in-cell codes aimed at modeling the laser-plasma dynamics with multiple ionization of atoms by strong external and plasma fields \cite{derouillat_cpc2018,mironov_pre25}.

\section*{Acknowledgment}
The authors are thankful to M.V. Frolov, A.V. Meremyanin, G.G. Paulus, and O.I. Tolstikhin for their interest to the work and for valuable discussions. DIT and SVP acknowledge support of the the Russian Science Foundation through Grant No.25-22-00308.

\appendix
\section{Modeling sequential ionization in the long and short pulse regimes by rate equations}\label{ap:A}

\begin{figure}[h!]
\includegraphics[width=0.85\linewidth]{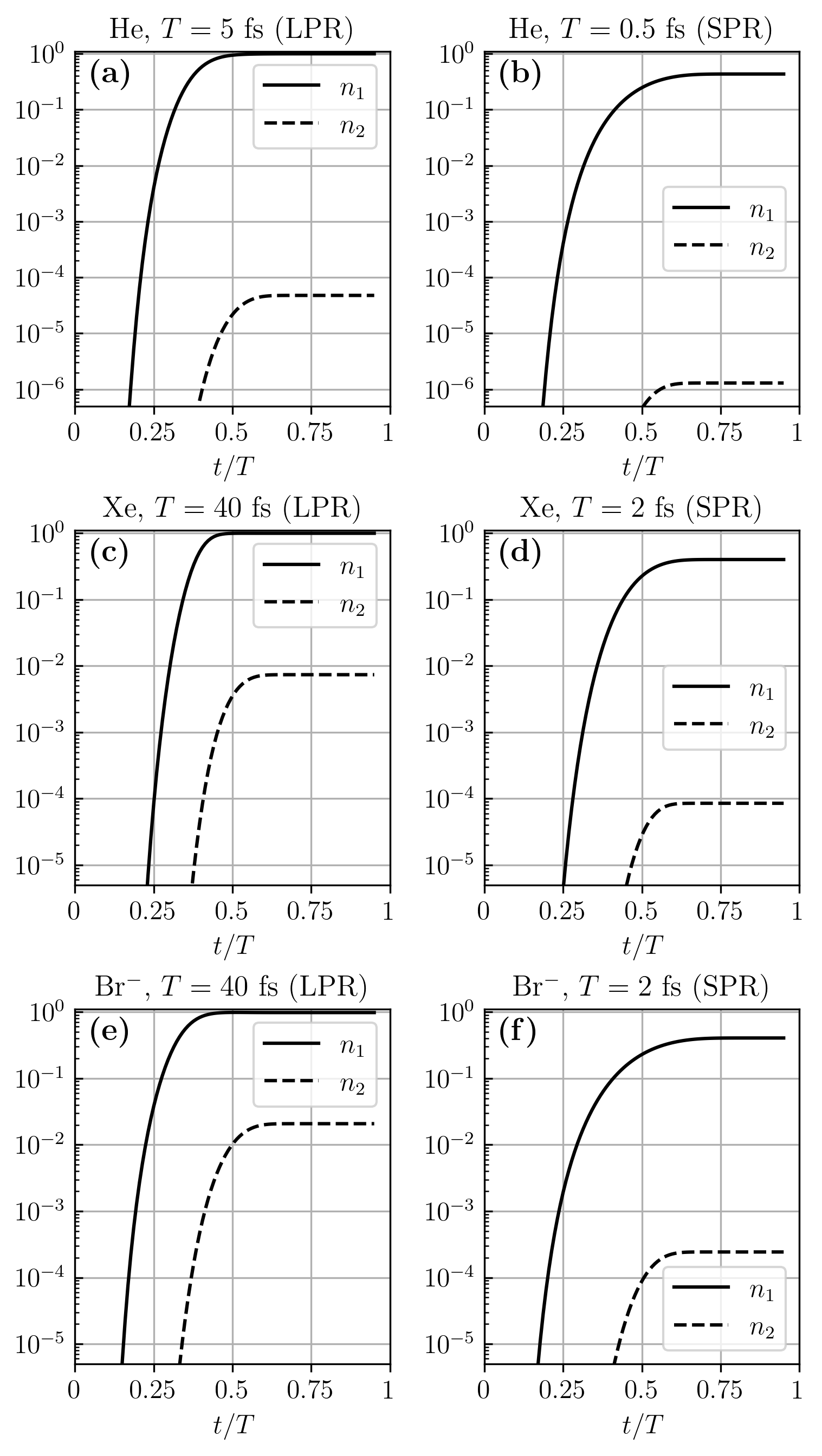}
\caption{
Time-dependent populations of singly (solid black lines) and doubly (dashed black lines) ionized atoms in the LPR (left column) and SPR (right column) for He, Xe and $\text{Br}^-$ calculated by solving numerically Eqs.(\ref{rate_eq}). 
}

\label{fig:concentrations}
\end{figure}

To estimate the pulse duration, which determines the border between the long pulse regime (LPR) and the short pulse regime (SPR)  for He, Xe and $\text{Br}^-$ , we solve numerically the system of rate equations based on the single active electron approximation:
\begin{equation} \label{rate_eq}
\begin{cases}
\dot{n}_0 = -w_1 n_0  \\
\dot{n}_1 = w_1 n_0 - w_2 n_1  \\
\dot{n}_2 = w_2 n_1~,~~n_0+n_1+n_2=1~,~~n_0(0)=1,  \\
\end{cases}
\end{equation}
Here $n_0$, $n_1$ and $n_2$ are the populations  of the seeding species (He, Xe or $\text{Br}^-$), singly ionized species ($\text{He}^+$, $\text{Xe}^+$ or $\text{Br}$) and doubly ionized species ($\text{He}^{2+}$, $\text{Xe}^{2+}$ or $\text{Br}^+$), correspondingly.
The probabilities per unit time (rates) of ionization $w_{1,2}$ can be calculated by different methods, although in any case, the SAE and the sequential process are assumed.
For the actual calculation whose results are shown in Fig.(\ref{fig:concentrations}) we use the semiclassical quasi-static formulas for the rate of ionization derived by Smirnov and Chibisov (SC) and Perelomov, Popov and Terent'yev (PPT)  \cite{smirnov_jetp66,popov_jetp66,perelomov_jetp67,popov-usp04}.
This rate reads
\begin{equation}
    w = 2^{2\nu+1}C^2 B_{lm} I_p F^{1+|m|-2\nu}(t) \exp\left( -\frac{2}{3F(t)} \right)~.
\label{PPT}
\end{equation}
Here
\begin{equation}
  F(t) = \frac{ E(t)}{(\rm 2 I_p)^{\rm 3/2}}  
  \label{F}
\end{equation}
is the reduced electric field scaled by the characteristic field $E_{\rm ch}=(2I_p)^{3/2}$ of the atomic level.
This notation must not be confused with the function defined by Eq.(\ref{Fx1x2}) in the main text.
The effective principal quantum number
\begin{equation}
    \nu = \frac{Z}{\sqrt{2I_p}}
    \label{nu}
\end{equation}
depends on the residual charge $Z$ of the atomic core,  which is $Z=0$ for negative ions and $Z=1$ for neutral atoms.
Next,
\begin{equation}
    C^2 = \dfrac{A^2}{4(2I_p)^{\nu+1/2}}
    \label{C}
\end{equation}
is the squared asymptotic coefficient of the bound state single-electron wave function, the factor $A$ was taken from \cite{radzig}, and
\begin{equation}
    B_{lm} = \frac{(2l+1)(l+|m|)!}{2^{2|m|}|m|!(l-|m|)!}
    \label{B}
\end{equation}
reflects the angular structure of this wave function.
For details of the derivation of the rate formula (\ref{PPT}) we send the reader to the reviews \cite{popov-usp04,poprz_jpb14} and references therein.

The saturation time $t_s$ is determined such that the population $n_1$ of singly ionized atoms reaches 0.8 at the field maximum.
We set $l=0,m=0$ for He and $l=1,m=0$ for Xe and ${\rm Br}^-$.
Numerical analysis of (\ref{rate_eq}) gives the results for $t_s$ presented in Table.\ref{tab:ts}. 
For Xe and $\text{Br}^-$ $t_s\approx 12$ fs,  while for He $t_s\approx 3$ fs.
Note that our \textit{ab initio} TDSE calculations are performed in the reduced dimensionality and correspond to model atomic systems with $l=0$. 
If we set $l=0$ for Xe and $\text{Br}^-$, the saturation time appears three times longer, $t_s \approx 36$ fs (this is obvious from (\ref{B})). 
Thus, the saturation times obtained from rate equations give only a rough estimation for pulse duration, below which the sequential ionization model breaks down. 
For $T<12$ fs the SPR is achieved for Xe and $\text{Br}^-$ regardless of whether $l=0$ or $l=1$.

\begin{table}[h]
    \centering
    \caption{ Saturation times for He, Xe and ${\rm Br}^-$ estimated using the SC/PPT ionization rates.}
    \label{tab:ts}
    \begin{tabular}{|l|c|c|c|}
        \hline
         & $E_0$ (a.u.) & $t_s$ (fs) & $l$ \\
        \hline
        He    & 0.34 & 3 & 0\\
        Xe    & 0.06 & 12& 1\\
        Br$^-$ & 0.035 & 12& 1\\
        \hline
    \end{tabular}
\end{table}

Fig.\ref{fig:concentrations} illustrates the difference between the LPR and the SPR for He, Xe and ${\rm Br}^-$ in terms of the time evolution of the concentrations. 
In the LPR, the outer level appears fully depleted before the ionization probability becomes significant for the inner electron.
Instead, in the SPR ionization of the inner orbital begins while the outer one is not yet entirely emptied.
Note that in the SPR, the population $n_2$ in the plateau domain appears 1.5-2 orders of magnitude lower than that in the LPR.
However, this suppression is balanced by the non-vanishing concentration $n_0$ of the seeding species, which makes the influence of the e-e interaction on the DI yield so significant as it follows from the plots of Figs.1,2.

Although the rates (\ref{PPT}) are routinely used in calculations of atomic ionization in strong low-frequency fields, one has to point out that there are two significant constraints limiting the applicability of (\ref{PPT}) in quantitative treatments of the ionization dynamics.
\begin{enumerate}
    \item Validity for ionization in short pulses. Originally, the rates were derived for the monochromatic field assuming a sufficiently long pulse, which is rather opposite to the case we consider in the paper.
    In short pulses, where the frequency $\omega$ is not a rigorously defined quantity, the definition of the tunneling ionization regime through the smallness of the Keldysh $\gamma$-parameter becomes ambiguous.
    This issue has been considered by Popov \cite{popov-jetp01} who showed that in short pulses (including unipolar pulses as we used in our calculations) the definition of the Keldysh parameter can be naturally generalized if one defines the frequency such that it makes the argument of the pulse-shape function dimensionless without any scaling factor.
    In our case, the field determined by Eq.(\ref{E(t)}) is characterized by the effective frequency $\omega=\pi/T$.
    
    This allows us to estimate the Keldysh parameters for the outer ($\gamma_1$) and the inner electron ($\gamma_2$) for the shortest pulses considered in our calculations, as summarized in Table~V.
    These estimates show that for such pulses either the tunneling regime ($\gamma<1$) or the intermediate regime $\gamma\simeq 1$ is realized.
    At longer pulse durations the Keldysh parameters become even smaller, corresponding to the deep tunneling regime.
    Thus, as far as the pulse duration is concerned, the use of the PPT rates is reasonably justified.
    
    \begin{table}[h]
    \centering
    \caption{ Keldysh parameters for the outer and the inner electron in He, Xe and ${\rm Br}^-$ calculated from (\ref{gamma}) for the shortest pulses considered in our calculations, viz. 0.5 fs for He and 2 fs for Xe and $\text{Br}^-$.}
    \label{tab:Keldysh_parameters}
    \begin{tabular}{|l|c|c|c|c|}
        \hline
         & $E_0$ (a.u.) & $T$ fs & $\gamma_1$ & $\gamma_2$ \\
        \hline
        He    & 0.34 & 0.5 & $0.60$ & $0.89$\\
        Xe    & 0.06 & 2& $0.60$ & $0.79$\\
        Br$^-$ & 0.035 & 2 & $0.54$ & $1.01$\\
        \hline
    \end{tabular}
\end{table}

    \item Another significant restriction of the SC and PPT rates based on the semi-classical approximation of quantum mechanics stems from the fact that this approximation becomes invalidated with field amplitude growing, for the barrier the electron tunnels through gets too short.
    Quantitatively, the border of the semi-classical approximation is predominantly determined by the value of the reduced field (\ref{F}), which has to be below $\approx 0.1$, see e.g. the review \cite{popov-usp04} and the literature quoted there.
    For higher fields, the PPT formula considerably overestimates the rate, and the latter can be empirically corrected.
    An efficient example of such correction was given in \cite{tong_jpb05}, although there are alternative ways known to extend the semi-classical results into the domain of higher fields, see e.g. \cite{bauer_pra1999,golovanov2020formula}.
    For our purposes, it is sufficient to note that the reduced field is close to this threshold or even exceeds it: $F=0.22$ for the outer electron at the field maximum, and $F=0.05$ for the inner electron. 
    Thus, the application of the PPT rates is perfectly justified in the case of sequential ionization of the inner electron, but remains questionable for the description of detachment of the first. 
    The situation is not that crucial, because the outer electron typically leaves the atom before the field maximum.
    However, the arguments presented above show that the semiclassical ionization rates can only be used for a qualitative analysis.
\end{enumerate}

Summarizing, the numerical modeling of the populations of the two electron levels is instructive for a qualitative estimation of the pulse duration, which separates the LPR from the SPR.
However, quantitative accuracy of such modeling  may not be expected.
This is the reason why, for our quantitative analysis, we compare numerical TDSE results with and without e-e interaction.

\bibliography{lit}

\begin{thebibliography}{61}%
\makeatletter
\providecommand \@ifxundefined [1]{%
 \@ifx{#1\undefined}
}%
\providecommand \@ifnum [1]{%
 \ifnum #1\expandafter \@firstoftwo
 \else \expandafter \@secondoftwo
 \fi
}%
\providecommand \@ifx [1]{%
 \ifx #1\expandafter \@firstoftwo
 \else \expandafter \@secondoftwo
 \fi
}%
\providecommand \natexlab [1]{#1}%
\providecommand \enquote  [1]{``#1''}%
\providecommand \bibnamefont  [1]{#1}%
\providecommand \bibfnamefont [1]{#1}%
\providecommand \citenamefont [1]{#1}%
\providecommand \href@noop [0]{\@secondoftwo}%
\providecommand \href [0]{\begingroup \@sanitize@url \@href}%
\providecommand \@href[1]{\@@startlink{#1}\@@href}%
\providecommand \@@href[1]{\endgroup#1\@@endlink}%
\providecommand \@sanitize@url [0]{\catcode `\\12\catcode `\$12\catcode
  `\&12\catcode `\#12\catcode `\^12\catcode `\_12\catcode `\%12\relax}%
\providecommand \@@startlink[1]{}%
\providecommand \@@endlink[0]{}%
\providecommand \url  [0]{\begingroup\@sanitize@url \@url }%
\providecommand \@url [1]{\endgroup\@href {#1}{\urlprefix }}%
\providecommand \urlprefix  [0]{URL }%
\providecommand \Eprint [0]{\href }%
\providecommand \doibase [0]{http://dx.doi.org/}%
\providecommand \selectlanguage [0]{\@gobble}%
\providecommand \bibinfo  [0]{\@secondoftwo}%
\providecommand \bibfield  [0]{\@secondoftwo}%
\providecommand \translation [1]{[#1]}%
\providecommand \BibitemOpen [0]{}%
\providecommand \bibitemStop [0]{}%
\providecommand \bibitemNoStop [0]{.\EOS\space}%
\providecommand \EOS [0]{\spacefactor3000\relax}%
\providecommand \BibitemShut  [1]{\csname bibitem#1\endcsname}%
\let\auto@bib@innerbib\@empty
\bibitem [{\citenamefont {Gamow}(1928)}]{gamow-28}%
  \BibitemOpen
  \bibfield  {author} {\bibinfo {author} {\bibfnamefont {G.}~\bibnamefont
  {Gamow}},\ }\href@noop {} {\bibfield  {journal} {\bibinfo  {journal} {Zc.
  Phys}\ }\textbf {\bibinfo {volume} {51}},\ \bibinfo {pages} {204} (\bibinfo
  {year} {1928})}\BibitemShut {NoStop}%
\bibitem [{\citenamefont {Gurney}\ and\ \citenamefont
  {Condon}(1929)}]{condon-pr29}%
  \BibitemOpen
  \bibfield  {author} {\bibinfo {author} {\bibfnamefont {R.~W.}\ \bibnamefont
  {Gurney}}\ and\ \bibinfo {author} {\bibfnamefont {E.~U.}\ \bibnamefont
  {Condon}},\ }\href@noop {} {\bibfield  {journal} {\bibinfo  {journal}
  {Physical Review}\ }\textbf {\bibinfo {volume} {33}},\ \bibinfo {pages} {127}
  (\bibinfo {year} {1929})}\BibitemShut {NoStop}%
\bibitem [{\citenamefont {Oppenheimer}(1928)}]{opp_pr28}%
  \BibitemOpen
  \bibfield  {author} {\bibinfo {author} {\bibfnamefont {J.~R.}\ \bibnamefont
  {Oppenheimer}},\ }\href {\doibase 10.1103/PhysRev.31.66} {\bibfield
  {journal} {\bibinfo  {journal} {Phys. Rev.}\ }\textbf {\bibinfo {volume}
  {31}},\ \bibinfo {pages} {66} (\bibinfo {year} {1928})}\BibitemShut {NoStop}%
\bibitem [{\citenamefont {Blatt}\ and\ \citenamefont
  {Weisskopf}(2012)}]{weisskopf-book}%
  \BibitemOpen
  \bibfield  {author} {\bibinfo {author} {\bibfnamefont {J.~M.}\ \bibnamefont
  {Blatt}}\ and\ \bibinfo {author} {\bibfnamefont {V.~F.}\ \bibnamefont
  {Weisskopf}},\ }\href@noop {} {\emph {\bibinfo {title} {Theoretical nuclear
  physics}}}\ (\bibinfo  {publisher} {Springer Science \& Business Media},\
  \bibinfo {year} {2012})\BibitemShut {NoStop}%
\bibitem [{\citenamefont {Fedorov}(1997)}]{fedorov-book}%
  \BibitemOpen
  \bibfield  {author} {\bibinfo {author} {\bibfnamefont {M.~V.}\ \bibnamefont
  {Fedorov}},\ }\href@noop {} {\emph {\bibinfo {title} {Atomic and free
  electrons in a strong light field}}}\ (\bibinfo  {publisher} {World
  Scientific},\ \bibinfo {year} {1997})\BibitemShut {NoStop}%
\bibitem [{\citenamefont {Popov}(2004)}]{popov-usp04}%
  \BibitemOpen
  \bibfield  {author} {\bibinfo {author} {\bibfnamefont {V.~S.}\ \bibnamefont
  {Popov}},\ }\href@noop {} {\bibfield  {journal} {\bibinfo  {journal}
  {Physics-Uspekhi}\ }\textbf {\bibinfo {volume} {47}},\ \bibinfo {pages} {855}
  (\bibinfo {year} {2004})}\BibitemShut {NoStop}%
\bibitem [{\citenamefont {Ankerhold}(2007)}]{ankerhold-07}%
  \BibitemOpen
  \bibfield  {author} {\bibinfo {author} {\bibfnamefont {J.}~\bibnamefont
  {Ankerhold}},\ }\href@noop {} {\emph {\bibinfo {title} {Quantum tunneling in
  complex systems: the semiclassical approach}}},\ Vol.\ \bibinfo {volume}
  {224}\ (\bibinfo  {publisher} {Springer},\ \bibinfo {year}
  {2007})\BibitemShut {NoStop}%
\bibitem [{\citenamefont {Krausz}\ and\ \citenamefont
  {Ivanov}(2009)}]{ivanov-rmp09}%
  \BibitemOpen
  \bibfield  {author} {\bibinfo {author} {\bibfnamefont {F.}~\bibnamefont
  {Krausz}}\ and\ \bibinfo {author} {\bibfnamefont {M.}~\bibnamefont
  {Ivanov}},\ }\href@noop {} {\bibfield  {journal} {\bibinfo  {journal} {Rev.
  Mod. Phys.}\ }\textbf {\bibinfo {volume} {81}},\ \bibinfo {pages} {163}
  (\bibinfo {year} {2009})}\BibitemShut {NoStop}%
\bibitem [{\citenamefont {Calegari}\ \emph {et~al.}(2016)\citenamefont
  {Calegari}, \citenamefont {Sansone}, \citenamefont {Stagira}, \citenamefont
  {Vozzi},\ and\ \citenamefont {Nisoli}}]{calegari-adv16}%
  \BibitemOpen
  \bibfield  {author} {\bibinfo {author} {\bibfnamefont {F.}~\bibnamefont
  {Calegari}}, \bibinfo {author} {\bibfnamefont {G.}~\bibnamefont {Sansone}},
  \bibinfo {author} {\bibfnamefont {S.}~\bibnamefont {Stagira}}, \bibinfo
  {author} {\bibfnamefont {C.}~\bibnamefont {Vozzi}}, \ and\ \bibinfo {author}
  {\bibfnamefont {M.}~\bibnamefont {Nisoli}},\ }\href@noop {} {\bibfield
  {journal} {\bibinfo  {journal} {Journal of Physics B: Atomic, Molecular and
  Optical Physics}\ }\textbf {\bibinfo {volume} {49}},\ \bibinfo {pages}
  {062001} (\bibinfo {year} {2016})}\BibitemShut {NoStop}%
\bibitem [{\citenamefont {Yue}\ and\ \citenamefont
  {Gaarde}(2022)}]{gaarde-josab22}%
  \BibitemOpen
  \bibfield  {author} {\bibinfo {author} {\bibfnamefont {L.}~\bibnamefont
  {Yue}}\ and\ \bibinfo {author} {\bibfnamefont {M.~B.}\ \bibnamefont
  {Gaarde}},\ }\href@noop {} {\bibfield  {journal} {\bibinfo  {journal} {JOSA
  B}\ }\textbf {\bibinfo {volume} {39}},\ \bibinfo {pages} {535} (\bibinfo
  {year} {2022})}\BibitemShut {NoStop}%
\bibitem [{\citenamefont {Ryabikin}\ \emph {et~al.}(2023)\citenamefont
  {Ryabikin}, \citenamefont {Emelin},\ and\ \citenamefont
  {Strelkov}}]{strelkov-usp23}%
  \BibitemOpen
  \bibfield  {author} {\bibinfo {author} {\bibfnamefont {M.}~\bibnamefont
  {Ryabikin}}, \bibinfo {author} {\bibfnamefont {M.}~\bibnamefont {Emelin}}, \
  and\ \bibinfo {author} {\bibfnamefont {V.~V.}\ \bibnamefont {Strelkov}},\
  }\href@noop {} {\bibfield  {journal} {\bibinfo  {journal} {Uspekhi
  Fizicheskikh Nauk}\ }\textbf {\bibinfo {volume} {193}},\ \bibinfo {pages}
  {382} (\bibinfo {year} {2023})}\BibitemShut {NoStop}%
\bibitem [{\citenamefont {Agostini}(2024)}]{agostini_nobel24}%
  \BibitemOpen
  \bibfield  {author} {\bibinfo {author} {\bibfnamefont {P.}~\bibnamefont
  {Agostini}},\ }\href {\doibase 10.1103/RevModPhys.96.030501} {\bibfield
  {journal} {\bibinfo  {journal} {Rev. Mod. Phys.}\ }\textbf {\bibinfo {volume}
  {96}},\ \bibinfo {pages} {030501} (\bibinfo {year} {2024})}\BibitemShut
  {NoStop}%
\bibitem [{\citenamefont {L'Huillier}(2024)}]{huillier_nobel24}%
  \BibitemOpen
  \bibfield  {author} {\bibinfo {author} {\bibfnamefont {A.}~\bibnamefont
  {L'Huillier}},\ }\href {\doibase 10.1103/RevModPhys.96.030503} {\bibfield
  {journal} {\bibinfo  {journal} {Rev. Mod. Phys.}\ }\textbf {\bibinfo {volume}
  {96}},\ \bibinfo {pages} {030503} (\bibinfo {year} {2024})}\BibitemShut
  {NoStop}%
\bibitem [{\citenamefont {Krausz}(2024)}]{krausz_nobel24}%
  \BibitemOpen
  \bibfield  {author} {\bibinfo {author} {\bibfnamefont {F.}~\bibnamefont
  {Krausz}},\ }\href {\doibase 10.1103/RevModPhys.96.030502} {\bibfield
  {journal} {\bibinfo  {journal} {Rev. Mod. Phys.}\ }\textbf {\bibinfo {volume}
  {96}},\ \bibinfo {pages} {030502} (\bibinfo {year} {2024})}\BibitemShut
  {NoStop}%
\bibitem [{\citenamefont {Keldysh}(1965)}]{keldysh-jetp65}%
  \BibitemOpen
  \bibfield  {author} {\bibinfo {author} {\bibfnamefont {L.}~\bibnamefont
  {Keldysh}},\ }\href@noop {} {\bibfield  {journal} {\bibinfo  {journal}
  {Soviet Phys. JETP}\ }\textbf {\bibinfo {volume} {20}} (\bibinfo {year}
  {1965})}\BibitemShut {NoStop}%
\bibitem [{\citenamefont {Voronov}\ and\ \citenamefont
  {Delone}(1966)}]{delone-jetp66}%
  \BibitemOpen
  \bibfield  {author} {\bibinfo {author} {\bibfnamefont {G.}~\bibnamefont
  {Voronov}}\ and\ \bibinfo {author} {\bibfnamefont {N.}~\bibnamefont
  {Delone}},\ }\href@noop {} {\bibfield  {journal} {\bibinfo  {journal} {Sov.
  Phys. JETP}\ }\textbf {\bibinfo {volume} {23}},\ \bibinfo {pages} {54}
  (\bibinfo {year} {1966})}\BibitemShut {NoStop}%
\bibitem [{\citenamefont {Agostini}\ \emph {et~al.}(1979)\citenamefont
  {Agostini}, \citenamefont {Fabre}, \citenamefont {Mainfray}, \citenamefont
  {Petite},\ and\ \citenamefont {Rahman}}]{agostini-prl79}%
  \BibitemOpen
  \bibfield  {author} {\bibinfo {author} {\bibfnamefont {P.}~\bibnamefont
  {Agostini}}, \bibinfo {author} {\bibfnamefont {F.}~\bibnamefont {Fabre}},
  \bibinfo {author} {\bibfnamefont {G.}~\bibnamefont {Mainfray}}, \bibinfo
  {author} {\bibfnamefont {G.}~\bibnamefont {Petite}}, \ and\ \bibinfo {author}
  {\bibfnamefont {N.~K.}\ \bibnamefont {Rahman}},\ }\href@noop {} {\bibfield
  {journal} {\bibinfo  {journal} {Phys. Rev. Lett.}\ }\textbf {\bibinfo
  {volume} {42}},\ \bibinfo {pages} {1127} (\bibinfo {year}
  {1979})}\BibitemShut {NoStop}%
\bibitem [{\citenamefont {Kuchiev}(1987)}]{kuchiev-jetpl87}%
  \BibitemOpen
  \bibfield  {author} {\bibinfo {author} {\bibfnamefont {M.~Y.}\ \bibnamefont
  {Kuchiev}},\ }\href@noop {} {\bibfield  {journal} {\bibinfo  {journal} {JETP
  Lett}\ }\textbf {\bibinfo {volume} {45}},\ \bibinfo {pages} {404} (\bibinfo
  {year} {1987})}\BibitemShut {NoStop}%
\bibitem [{\citenamefont {Corkum}(1993)}]{corkum-prl93}%
  \BibitemOpen
  \bibfield  {author} {\bibinfo {author} {\bibfnamefont {P.~B.}\ \bibnamefont
  {Corkum}},\ }\href@noop {} {\bibfield  {journal} {\bibinfo  {journal} {Phys.
  Rev. Lett.}\ }\textbf {\bibinfo {volume} {71}},\ \bibinfo {pages} {1994}
  (\bibinfo {year} {1993})}\BibitemShut {NoStop}%
\bibitem [{\citenamefont {Paulus}\ \emph {et~al.}(1994)\citenamefont {Paulus},
  \citenamefont {Nicklich}, \citenamefont {Xu}, \citenamefont {Lambropoulos},\
  and\ \citenamefont {Walther}}]{paulus-prl94}%
  \BibitemOpen
  \bibfield  {author} {\bibinfo {author} {\bibfnamefont {G.}~\bibnamefont
  {Paulus}}, \bibinfo {author} {\bibfnamefont {W.}~\bibnamefont {Nicklich}},
  \bibinfo {author} {\bibfnamefont {H.}~\bibnamefont {Xu}}, \bibinfo {author}
  {\bibfnamefont {P.}~\bibnamefont {Lambropoulos}}, \ and\ \bibinfo {author}
  {\bibfnamefont {H.}~\bibnamefont {Walther}},\ }\href@noop {} {\bibfield
  {journal} {\bibinfo  {journal} {Phys. Rev. Lett.}\ }\textbf {\bibinfo
  {volume} {72}},\ \bibinfo {pages} {2851} (\bibinfo {year}
  {1994})}\BibitemShut {NoStop}%
\bibitem [{\citenamefont {Becker}\ \emph {et~al.}(2012)\citenamefont {Becker},
  \citenamefont {Liu}, \citenamefont {Ho},\ and\ \citenamefont
  {Eberly}}]{becker-rmp12}%
  \BibitemOpen
  \bibfield  {author} {\bibinfo {author} {\bibfnamefont {W.}~\bibnamefont
  {Becker}}, \bibinfo {author} {\bibfnamefont {X.}~\bibnamefont {Liu}},
  \bibinfo {author} {\bibfnamefont {P.~J.}\ \bibnamefont {Ho}}, \ and\ \bibinfo
  {author} {\bibfnamefont {J.~H.}\ \bibnamefont {Eberly}},\ }\href@noop {}
  {\bibfield  {journal} {\bibinfo  {journal} {Rev. Mod. Phys.}\ }\textbf
  {\bibinfo {volume} {84}},\ \bibinfo {pages} {1011} (\bibinfo {year}
  {2012})}\BibitemShut {NoStop}%
\bibitem [{\citenamefont {Dimauro}\ and\ \citenamefont
  {Agostini}(1995)}]{agostini-rev95}%
  \BibitemOpen
  \bibfield  {author} {\bibinfo {author} {\bibfnamefont {L.}~\bibnamefont
  {Dimauro}}\ and\ \bibinfo {author} {\bibfnamefont {P.}~\bibnamefont
  {Agostini}},\ }in\ \href@noop {} {\emph {\bibinfo {booktitle} {Advances in
  Atomic, Molecular, and Optical Physics}}},\ Vol.~\bibinfo {volume} {35}\
  (\bibinfo  {publisher} {Elsevier},\ \bibinfo {year} {1995})\ pp.\ \bibinfo
  {pages} {79--120}\BibitemShut {NoStop}%
\bibitem [{\citenamefont {Zon}(1999)}]{zon-jetp99}%
  \BibitemOpen
  \bibfield  {author} {\bibinfo {author} {\bibfnamefont {B.}~\bibnamefont
  {Zon}},\ }\href@noop {} {\bibfield  {journal} {\bibinfo  {journal} {Journal
  of Experimental and Theoretical Physics}\ }\textbf {\bibinfo {volume} {89}},\
  \bibinfo {pages} {219} (\bibinfo {year} {1999})}\BibitemShut {NoStop}%
\bibitem [{\citenamefont {Eichmann}\ \emph {et~al.}(2000)\citenamefont
  {Eichmann}, \citenamefont {D\"orr}, \citenamefont {Maeda}, \citenamefont
  {Becker},\ and\ \citenamefont {Sandner}}]{becker-prl00}%
  \BibitemOpen
  \bibfield  {author} {\bibinfo {author} {\bibfnamefont {U.}~\bibnamefont
  {Eichmann}}, \bibinfo {author} {\bibfnamefont {M.}~\bibnamefont {D\"orr}},
  \bibinfo {author} {\bibfnamefont {H.}~\bibnamefont {Maeda}}, \bibinfo
  {author} {\bibfnamefont {W.}~\bibnamefont {Becker}}, \ and\ \bibinfo {author}
  {\bibfnamefont {W.}~\bibnamefont {Sandner}},\ }\href {\doibase
  10.1103/PhysRevLett.84.3550} {\bibfield  {journal} {\bibinfo  {journal}
  {Phys. Rev. Lett.}\ }\textbf {\bibinfo {volume} {84}},\ \bibinfo {pages}
  {3550} (\bibinfo {year} {2000})}\BibitemShut {NoStop}%
\bibitem [{\citenamefont {Popruzhenko}\ and\ \citenamefont
  {Tyurin}(2023)}]{tyurin-leb23}%
  \BibitemOpen
  \bibfield  {author} {\bibinfo {author} {\bibfnamefont {S.}~\bibnamefont
  {Popruzhenko}}\ and\ \bibinfo {author} {\bibfnamefont {D.}~\bibnamefont
  {Tyurin}},\ }\href@noop {} {\bibfield  {journal} {\bibinfo  {journal}
  {Bulletin of the Lebedev Physics Institute}\ }\textbf {\bibinfo {volume}
  {50}},\ \bibinfo {pages} {S922} (\bibinfo {year} {2023})}\BibitemShut
  {NoStop}%
\bibitem [{\citenamefont {Lein}\ \emph {et~al.}(2000)\citenamefont {Lein},
  \citenamefont {Gross},\ and\ \citenamefont {Engel}}]{lein_prl00}%
  \BibitemOpen
  \bibfield  {author} {\bibinfo {author} {\bibfnamefont {M.}~\bibnamefont
  {Lein}}, \bibinfo {author} {\bibfnamefont {E.~K.}\ \bibnamefont {Gross}}, \
  and\ \bibinfo {author} {\bibfnamefont {V.}~\bibnamefont {Engel}},\
  }\href@noop {} {\bibfield  {journal} {\bibinfo  {journal} {Phys. Rev. Lett.}\
  }\textbf {\bibinfo {volume} {85}},\ \bibinfo {pages} {4707} (\bibinfo {year}
  {2000})}\BibitemShut {NoStop}%
\bibitem [{\citenamefont {Ho}\ \emph {et~al.}(2005)\citenamefont {Ho},
  \citenamefont {Panfili}, \citenamefont {Haan},\ and\ \citenamefont
  {Eberly}}]{eberly_prl05}%
  \BibitemOpen
  \bibfield  {author} {\bibinfo {author} {\bibfnamefont {P.~J.}\ \bibnamefont
  {Ho}}, \bibinfo {author} {\bibfnamefont {R.}~\bibnamefont {Panfili}},
  \bibinfo {author} {\bibfnamefont {S.~L.}\ \bibnamefont {Haan}}, \ and\
  \bibinfo {author} {\bibfnamefont {J.}~\bibnamefont {Eberly}},\ }\href@noop {}
  {\bibfield  {journal} {\bibinfo  {journal} {Physical review letters}\
  }\textbf {\bibinfo {volume} {94}},\ \bibinfo {pages} {093002} (\bibinfo
  {year} {2005})}\BibitemShut {NoStop}%
\bibitem [{\citenamefont {Mauger}\ \emph {et~al.}(2012)\citenamefont {Mauger},
  \citenamefont {Kamor}, \citenamefont {Chandre},\ and\ \citenamefont
  {Uzer}}]{mauger_prl12}%
  \BibitemOpen
  \bibfield  {author} {\bibinfo {author} {\bibfnamefont {F.}~\bibnamefont
  {Mauger}}, \bibinfo {author} {\bibfnamefont {A.}~\bibnamefont {Kamor}},
  \bibinfo {author} {\bibfnamefont {C.}~\bibnamefont {Chandre}}, \ and\
  \bibinfo {author} {\bibfnamefont {T.}~\bibnamefont {Uzer}},\ }\href@noop {}
  {\bibfield  {journal} {\bibinfo  {journal} {Physical Review Letters}\
  }\textbf {\bibinfo {volume} {108}},\ \bibinfo {pages} {063001} (\bibinfo
  {year} {2012})}\BibitemShut {NoStop}%
\bibitem [{\citenamefont {Huang}\ \emph {et~al.}(2013)\citenamefont {Huang},
  \citenamefont {Zhou}, \citenamefont {Zhang},\ and\ \citenamefont
  {Lu}}]{huang_oe13}%
  \BibitemOpen
  \bibfield  {author} {\bibinfo {author} {\bibfnamefont {C.}~\bibnamefont
  {Huang}}, \bibinfo {author} {\bibfnamefont {Y.}~\bibnamefont {Zhou}},
  \bibinfo {author} {\bibfnamefont {Q.}~\bibnamefont {Zhang}}, \ and\ \bibinfo
  {author} {\bibfnamefont {P.}~\bibnamefont {Lu}},\ }\href@noop {} {\bibfield
  {journal} {\bibinfo  {journal} {Optics Express}\ }\textbf {\bibinfo {volume}
  {21}},\ \bibinfo {pages} {11382} (\bibinfo {year} {2013})}\BibitemShut
  {NoStop}%
\bibitem [{\citenamefont {Kopold}\ \emph {et~al.}(2000)\citenamefont {Kopold},
  \citenamefont {Becker}, \citenamefont {Rottke},\ and\ \citenamefont
  {Sandner}}]{kopold_prl00}%
  \BibitemOpen
  \bibfield  {author} {\bibinfo {author} {\bibfnamefont {R.}~\bibnamefont
  {Kopold}}, \bibinfo {author} {\bibfnamefont {W.}~\bibnamefont {Becker}},
  \bibinfo {author} {\bibfnamefont {H.}~\bibnamefont {Rottke}}, \ and\ \bibinfo
  {author} {\bibfnamefont {W.}~\bibnamefont {Sandner}},\ }\href@noop {}
  {\bibfield  {journal} {\bibinfo  {journal} {Physical review letters}\
  }\textbf {\bibinfo {volume} {85}},\ \bibinfo {pages} {3781} (\bibinfo {year}
  {2000})}\BibitemShut {NoStop}%
\bibitem [{\citenamefont {Rudenko}\ \emph {et~al.}(2004)\citenamefont
  {Rudenko}, \citenamefont {Zrost}, \citenamefont {Feuerstein}, \citenamefont
  {de~Jesus}, \citenamefont {Schr\"oter}, \citenamefont {Moshammer},\ and\
  \citenamefont {Ullrich}}]{rudenko_prl04}%
  \BibitemOpen
  \bibfield  {author} {\bibinfo {author} {\bibfnamefont {A.}~\bibnamefont
  {Rudenko}}, \bibinfo {author} {\bibfnamefont {K.}~\bibnamefont {Zrost}},
  \bibinfo {author} {\bibfnamefont {B.}~\bibnamefont {Feuerstein}}, \bibinfo
  {author} {\bibfnamefont {V.~L.~B.}\ \bibnamefont {de~Jesus}}, \bibinfo
  {author} {\bibfnamefont {C.~D.}\ \bibnamefont {Schr\"oter}}, \bibinfo
  {author} {\bibfnamefont {R.}~\bibnamefont {Moshammer}}, \ and\ \bibinfo
  {author} {\bibfnamefont {J.}~\bibnamefont {Ullrich}},\ }\href {\doibase
  10.1103/PhysRevLett.93.253001} {\bibfield  {journal} {\bibinfo  {journal}
  {Phys. Rev. Lett.}\ }\textbf {\bibinfo {volume} {93}},\ \bibinfo {pages}
  {253001} (\bibinfo {year} {2004})}\BibitemShut {NoStop}%
\bibitem [{\citenamefont {Emmanouilidou}\ and\ \citenamefont
  {Staudte}(2009)}]{emma_pra09}%
  \BibitemOpen
  \bibfield  {author} {\bibinfo {author} {\bibfnamefont {A.}~\bibnamefont
  {Emmanouilidou}}\ and\ \bibinfo {author} {\bibfnamefont {A.}~\bibnamefont
  {Staudte}},\ }\href@noop {} {\bibfield  {journal} {\bibinfo  {journal}
  {Physical Review A—Atomic, Molecular, and Optical Physics}\ }\textbf
  {\bibinfo {volume} {80}},\ \bibinfo {pages} {053415} (\bibinfo {year}
  {2009})}\BibitemShut {NoStop}%
\bibitem [{\citenamefont {Klaiber}\ \emph {et~al.}(2007)\citenamefont
  {Klaiber}, \citenamefont {Hatsagortsyan},\ and\ \citenamefont
  {Keitel}}]{keitel-pra07}%
  \BibitemOpen
  \bibfield  {author} {\bibinfo {author} {\bibfnamefont {M.}~\bibnamefont
  {Klaiber}}, \bibinfo {author} {\bibfnamefont {K.~Z.}\ \bibnamefont
  {Hatsagortsyan}}, \ and\ \bibinfo {author} {\bibfnamefont {C.~H.}\
  \bibnamefont {Keitel}},\ }\href@noop {} {\bibfield  {journal} {\bibinfo
  {journal} {Physical Review A—Atomic, Molecular, and Optical Physics}\
  }\textbf {\bibinfo {volume} {75}},\ \bibinfo {pages} {063413} (\bibinfo
  {year} {2007})}\BibitemShut {NoStop}%
\bibitem [{\citenamefont {Shvetsov-Shilovski}\ \emph
  {et~al.}(2008)\citenamefont {Shvetsov-Shilovski}, \citenamefont
  {Goreslavski}, \citenamefont {Popruzhenko},\ and\ \citenamefont
  {Becker}}]{kolya-pra08}%
  \BibitemOpen
  \bibfield  {author} {\bibinfo {author} {\bibfnamefont {N.}~\bibnamefont
  {Shvetsov-Shilovski}}, \bibinfo {author} {\bibfnamefont {S.}~\bibnamefont
  {Goreslavski}}, \bibinfo {author} {\bibfnamefont {S.}~\bibnamefont
  {Popruzhenko}}, \ and\ \bibinfo {author} {\bibfnamefont {W.}~\bibnamefont
  {Becker}},\ }\href@noop {} {\bibfield  {journal} {\bibinfo  {journal}
  {Physical Review A—Atomic, Molecular, and Optical Physics}\ }\textbf
  {\bibinfo {volume} {77}},\ \bibinfo {pages} {063405} (\bibinfo {year}
  {2008})}\BibitemShut {NoStop}%
\bibitem [{\citenamefont {K{\"u}bel}\ \emph {et~al.}(2014)\citenamefont
  {K{\"u}bel}, \citenamefont {Betsch}, \citenamefont {Kling}, \citenamefont
  {Alnaser}, \citenamefont {Schmidt}, \citenamefont {Kleineberg}, \citenamefont
  {Deng}, \citenamefont {Ben-Itzhak}, \citenamefont {Paulus}, \citenamefont
  {Pfeifer} \emph {et~al.}}]{kubel_njp14}%
  \BibitemOpen
  \bibfield  {author} {\bibinfo {author} {\bibfnamefont {M.}~\bibnamefont
  {K{\"u}bel}}, \bibinfo {author} {\bibfnamefont {K.}~\bibnamefont {Betsch}},
  \bibinfo {author} {\bibfnamefont {N.~G.}\ \bibnamefont {Kling}}, \bibinfo
  {author} {\bibfnamefont {A.}~\bibnamefont {Alnaser}}, \bibinfo {author}
  {\bibfnamefont {J.}~\bibnamefont {Schmidt}}, \bibinfo {author} {\bibfnamefont
  {U.}~\bibnamefont {Kleineberg}}, \bibinfo {author} {\bibfnamefont
  {Y.}~\bibnamefont {Deng}}, \bibinfo {author} {\bibfnamefont {I.}~\bibnamefont
  {Ben-Itzhak}}, \bibinfo {author} {\bibfnamefont {G.}~\bibnamefont {Paulus}},
  \bibinfo {author} {\bibfnamefont {T.}~\bibnamefont {Pfeifer}},  \emph
  {et~al.},\ }\href@noop {} {\bibfield  {journal} {\bibinfo  {journal} {New
  Journal of Physics}\ }\textbf {\bibinfo {volume} {16}},\ \bibinfo {pages}
  {033008} (\bibinfo {year} {2014})}\BibitemShut {NoStop}%
\bibitem [{\citenamefont {Klaiber}\ \emph {et~al.}(2022)\citenamefont
  {Klaiber}, \citenamefont {Lv}, \citenamefont {Hatsagortsyan},\ and\
  \citenamefont {Keitel}}]{keitel-pra22}%
  \BibitemOpen
  \bibfield  {author} {\bibinfo {author} {\bibfnamefont {M.}~\bibnamefont
  {Klaiber}}, \bibinfo {author} {\bibfnamefont {Q.}~\bibnamefont {Lv}},
  \bibinfo {author} {\bibfnamefont {K.~Z.}\ \bibnamefont {Hatsagortsyan}}, \
  and\ \bibinfo {author} {\bibfnamefont {C.~H.}\ \bibnamefont {Keitel}},\
  }\href@noop {} {\bibfield  {journal} {\bibinfo  {journal} {Physical Review
  A}\ }\textbf {\bibinfo {volume} {105}},\ \bibinfo {pages} {063109} (\bibinfo
  {year} {2022})}\BibitemShut {NoStop}%
\bibitem [{\citenamefont {Fittinghoff}\ \emph {et~al.}(1992)\citenamefont
  {Fittinghoff}, \citenamefont {Bolton}, \citenamefont {Chang},\ and\
  \citenamefont {Kulander}}]{kulander_prl92}%
  \BibitemOpen
  \bibfield  {author} {\bibinfo {author} {\bibfnamefont {D.~N.}\ \bibnamefont
  {Fittinghoff}}, \bibinfo {author} {\bibfnamefont {P.~R.}\ \bibnamefont
  {Bolton}}, \bibinfo {author} {\bibfnamefont {B.}~\bibnamefont {Chang}}, \
  and\ \bibinfo {author} {\bibfnamefont {K.~C.}\ \bibnamefont {Kulander}},\
  }\href@noop {} {\bibfield  {journal} {\bibinfo  {journal} {Physical review
  letters}\ }\textbf {\bibinfo {volume} {69}},\ \bibinfo {pages} {2642}
  (\bibinfo {year} {1992})}\BibitemShut {NoStop}%
\bibitem [{\citenamefont {Feuerstein}\ \emph {et~al.}(2001)\citenamefont
  {Feuerstein}, \citenamefont {Moshammer}, \citenamefont {Fischer},
  \citenamefont {Dorn}, \citenamefont {Schr{\"o}ter}, \citenamefont
  {Deipenwisch}, \citenamefont {Lopez-Urrutia}, \citenamefont {H{\"o}hr},
  \citenamefont {Neumayer}, \citenamefont {Ullrich} \emph
  {et~al.}}]{moshammer_prl01}%
  \BibitemOpen
  \bibfield  {author} {\bibinfo {author} {\bibfnamefont {B.}~\bibnamefont
  {Feuerstein}}, \bibinfo {author} {\bibfnamefont {R.}~\bibnamefont
  {Moshammer}}, \bibinfo {author} {\bibfnamefont {D.}~\bibnamefont {Fischer}},
  \bibinfo {author} {\bibfnamefont {A.}~\bibnamefont {Dorn}}, \bibinfo {author}
  {\bibfnamefont {C.~D.}\ \bibnamefont {Schr{\"o}ter}}, \bibinfo {author}
  {\bibfnamefont {J.}~\bibnamefont {Deipenwisch}}, \bibinfo {author}
  {\bibfnamefont {J.~C.}\ \bibnamefont {Lopez-Urrutia}}, \bibinfo {author}
  {\bibfnamefont {C.}~\bibnamefont {H{\"o}hr}}, \bibinfo {author}
  {\bibfnamefont {P.}~\bibnamefont {Neumayer}}, \bibinfo {author}
  {\bibfnamefont {J.}~\bibnamefont {Ullrich}},  \emph {et~al.},\ }\href@noop {}
  {\bibfield  {journal} {\bibinfo  {journal} {Physical review letters}\
  }\textbf {\bibinfo {volume} {87}},\ \bibinfo {pages} {043003} (\bibinfo
  {year} {2001})}\BibitemShut {NoStop}%
\bibitem [{\citenamefont {Hall}\ \emph {et~al.}(2009)\citenamefont {Hall},
  \citenamefont {Saad}, \citenamefont {Sen},\ and\ \citenamefont
  {Ciftci}}]{hall2009energies}%
  \BibitemOpen
  \bibfield  {author} {\bibinfo {author} {\bibfnamefont {R.~L.}\ \bibnamefont
  {Hall}}, \bibinfo {author} {\bibfnamefont {N.}~\bibnamefont {Saad}}, \bibinfo
  {author} {\bibfnamefont {K.}~\bibnamefont {Sen}}, \ and\ \bibinfo {author}
  {\bibfnamefont {H.}~\bibnamefont {Ciftci}},\ }\href@noop {} {\bibfield
  {journal} {\bibinfo  {journal} {Physical Review A—Atomic, Molecular, and
  Optical Physics}\ }\textbf {\bibinfo {volume} {80}},\ \bibinfo {pages}
  {032507} (\bibinfo {year} {2009})}\BibitemShut {NoStop}%
\bibitem [{\citenamefont {Fern{\'a}ndez}(1991)}]{fernandez1991analytical}%
  \BibitemOpen
  \bibfield  {author} {\bibinfo {author} {\bibfnamefont {F.}~\bibnamefont
  {Fern{\'a}ndez}},\ }\href@noop {} {\bibfield  {journal} {\bibinfo  {journal}
  {Journal of Physics A: Mathematical and General}\ }\textbf {\bibinfo {volume}
  {24}},\ \bibinfo {pages} {1351} (\bibinfo {year} {1991})}\BibitemShut
  {NoStop}%
\bibitem [{\citenamefont {Lima}\ and\ \citenamefont
  {Miranda}(1981)}]{lima1981atoms}%
  \BibitemOpen
  \bibfield  {author} {\bibinfo {author} {\bibfnamefont {C.}~\bibnamefont
  {Lima}}\ and\ \bibinfo {author} {\bibfnamefont {L.}~\bibnamefont {Miranda}},\
  }\href@noop {} {\bibfield  {journal} {\bibinfo  {journal} {Physical Review
  A}\ }\textbf {\bibinfo {volume} {23}},\ \bibinfo {pages} {3335} (\bibinfo
  {year} {1981})}\BibitemShut {NoStop}%
\bibitem [{\citenamefont {Mehta}\ and\ \citenamefont
  {Patil}(1978)}]{mehta1978bound}%
  \BibitemOpen
  \bibfield  {author} {\bibinfo {author} {\bibfnamefont {C.}~\bibnamefont
  {Mehta}}\ and\ \bibinfo {author} {\bibfnamefont {S.}~\bibnamefont {Patil}},\
  }\href@noop {} {\bibfield  {journal} {\bibinfo  {journal} {Physical Review
  A}\ }\textbf {\bibinfo {volume} {17}},\ \bibinfo {pages} {43} (\bibinfo
  {year} {1978})}\BibitemShut {NoStop}%
\bibitem [{\citenamefont {Silaev}\ \emph {et~al.}(2010)\citenamefont {Silaev},
  \citenamefont {Ryabikin},\ and\ \citenamefont
  {Vvedenskii}}]{silaev2010strong}%
  \BibitemOpen
  \bibfield  {author} {\bibinfo {author} {\bibfnamefont {A.}~\bibnamefont
  {Silaev}}, \bibinfo {author} {\bibfnamefont {M.~Y.}\ \bibnamefont
  {Ryabikin}}, \ and\ \bibinfo {author} {\bibfnamefont {N.}~\bibnamefont
  {Vvedenskii}},\ }\href@noop {} {\bibfield  {journal} {\bibinfo  {journal}
  {Physical Review A—Atomic, Molecular, and Optical Physics}\ }\textbf
  {\bibinfo {volume} {82}},\ \bibinfo {pages} {033416} (\bibinfo {year}
  {2010})}\BibitemShut {NoStop}%
\bibitem [{\citenamefont {Gebremedhin}\ and\ \citenamefont
  {Weatherford}(2014)}]{gebremedhin2014calculations}%
  \BibitemOpen
  \bibfield  {author} {\bibinfo {author} {\bibfnamefont {D.~H.}\ \bibnamefont
  {Gebremedhin}}\ and\ \bibinfo {author} {\bibfnamefont {C.~A.}\ \bibnamefont
  {Weatherford}},\ }\href@noop {} {\bibfield  {journal} {\bibinfo  {journal}
  {Physical Review E}\ }\textbf {\bibinfo {volume} {89}},\ \bibinfo {pages}
  {053319} (\bibinfo {year} {2014})}\BibitemShut {NoStop}%
\bibitem [{\citenamefont {Weideman}\ and\ \citenamefont
  {Herbst}(1986)}]{weideman1986split}%
  \BibitemOpen
  \bibfield  {author} {\bibinfo {author} {\bibfnamefont {J.~A.~C.}\
  \bibnamefont {Weideman}}\ and\ \bibinfo {author} {\bibfnamefont {B.~M.}\
  \bibnamefont {Herbst}},\ }\href@noop {} {\bibfield  {journal} {\bibinfo
  {journal} {SIAM Journal on Numerical Analysis}\ }\textbf {\bibinfo {volume}
  {23}},\ \bibinfo {pages} {485} (\bibinfo {year} {1986})}\BibitemShut
  {NoStop}%
\bibitem [{\citenamefont {Neuhasuer}\ and\ \citenamefont
  {Baer}(1989)}]{neuhasuer1989}%
  \BibitemOpen
  \bibfield  {author} {\bibinfo {author} {\bibfnamefont {D.}~\bibnamefont
  {Neuhasuer}}\ and\ \bibinfo {author} {\bibfnamefont {M.}~\bibnamefont
  {Baer}},\ }\href@noop {} {\bibfield  {journal} {\bibinfo  {journal} {The
  Journal of chemical physics}\ }\textbf {\bibinfo {volume} {90}},\ \bibinfo
  {pages} {4351} (\bibinfo {year} {1989})}\BibitemShut {NoStop}%
\bibitem [{\citenamefont {Scrinzi}(2012)}]{scrinzi2012}%
  \BibitemOpen
  \bibfield  {author} {\bibinfo {author} {\bibfnamefont {A.}~\bibnamefont
  {Scrinzi}},\ }\href@noop {} {\bibfield  {journal} {\bibinfo  {journal} {New
  Journal of Physics}\ }\textbf {\bibinfo {volume} {14}},\ \bibinfo {pages}
  {085008} (\bibinfo {year} {2012})}\BibitemShut {NoStop}%
\bibitem [{\citenamefont {Bader}\ \emph {et~al.}(2013)\citenamefont {Bader},
  \citenamefont {Blanes},\ and\ \citenamefont {Casas}}]{bader2013}%
  \BibitemOpen
  \bibfield  {author} {\bibinfo {author} {\bibfnamefont {P.}~\bibnamefont
  {Bader}}, \bibinfo {author} {\bibfnamefont {S.}~\bibnamefont {Blanes}}, \
  and\ \bibinfo {author} {\bibfnamefont {F.}~\bibnamefont {Casas}},\
  }\href@noop {} {\bibfield  {journal} {\bibinfo  {journal} {The Journal of
  chemical physics}\ }\textbf {\bibinfo {volume} {139}} (\bibinfo {year}
  {2013})}\BibitemShut {NoStop}%
\bibitem [{\citenamefont {Smirnov}\ and\ \citenamefont
  {Chibisov}(1966)}]{smirnov_jetp66}%
  \BibitemOpen
  \bibfield  {author} {\bibinfo {author} {\bibfnamefont {B.}~\bibnamefont
  {Smirnov}}\ and\ \bibinfo {author} {\bibfnamefont {M.}~\bibnamefont
  {Chibisov}},\ }\href@noop {} {\bibfield  {journal} {\bibinfo  {journal} {Sov.
  Phys. JETP}\ }\textbf {\bibinfo {volume} {22}},\ \bibinfo {pages} {23}
  (\bibinfo {year} {1966})}\BibitemShut {NoStop}%
\bibitem [{\citenamefont {Perelomov}\ \emph {et~al.}(1966)\citenamefont
  {Perelomov}, \citenamefont {Popov},\ and\ \citenamefont
  {Terent’ev}}]{popov_jetp66}%
  \BibitemOpen
  \bibfield  {author} {\bibinfo {author} {\bibfnamefont {A.}~\bibnamefont
  {Perelomov}}, \bibinfo {author} {\bibfnamefont {V.}~\bibnamefont {Popov}}, \
  and\ \bibinfo {author} {\bibfnamefont {M.}~\bibnamefont {Terent’ev}},\
  }\href@noop {} {\bibfield  {journal} {\bibinfo  {journal} {Zh. Eksp. Teor.
  Fiz}\ }\textbf {\bibinfo {volume} {50}},\ \bibinfo {pages} {1393} (\bibinfo
  {year} {1966})}\BibitemShut {NoStop}%
\bibitem [{\citenamefont {Perelomov}\ \emph {et~al.}(1967)\citenamefont
  {Perelomov}, \citenamefont {Popov},\ and\ \citenamefont
  {Terent’ev}}]{perelomov_jetp67}%
  \BibitemOpen
  \bibfield  {author} {\bibinfo {author} {\bibfnamefont {A.}~\bibnamefont
  {Perelomov}}, \bibinfo {author} {\bibfnamefont {V.}~\bibnamefont {Popov}}, \
  and\ \bibinfo {author} {\bibfnamefont {M.}~\bibnamefont {Terent’ev}},\
  }\href@noop {} {\bibfield  {journal} {\bibinfo  {journal} {Zh. Eksp. Teor.
  Fiz}\ }\textbf {\bibinfo {volume} {52}},\ \bibinfo {pages} {514} (\bibinfo
  {year} {1967})}\BibitemShut {NoStop}%
\bibitem [{\citenamefont {Popruzhenko}(2014)}]{poprz_jpb14}%
  \BibitemOpen
  \bibfield  {author} {\bibinfo {author} {\bibfnamefont {S.}~\bibnamefont
  {Popruzhenko}},\ }\href@noop {} {\bibfield  {journal} {\bibinfo  {journal}
  {Journal of Physics B: Atomic, Molecular and Optical Physics}\ }\textbf
  {\bibinfo {volume} {47}},\ \bibinfo {pages} {204001} (\bibinfo {year}
  {2014})}\BibitemShut {NoStop}%
\bibitem [{\citenamefont {Tyurin}\ \emph {et~al.}(2026)\citenamefont {Tyurin},
  \citenamefont {Strelkov},\ and\ \citenamefont
  {Popruzhenko}}]{tyurin_arxiv26}%
  \BibitemOpen
  \bibfield  {author} {\bibinfo {author} {\bibfnamefont {D.}~\bibnamefont
  {Tyurin}}, \bibinfo {author} {\bibfnamefont {V.}~\bibnamefont {Strelkov}}, \
  and\ \bibinfo {author} {\bibfnamefont {S.}~\bibnamefont {Popruzhenko}},\
  }\href@noop {} {\bibfield  {journal} {\bibinfo  {journal} {arXiv preprint
  arXiv:2608.03998}\ } (\bibinfo {year} {2026})}\BibitemShut {NoStop}%
\bibitem [{\citenamefont {Bart}\ \emph {et~al.}(2020)\citenamefont {Bart},
  \citenamefont {Schubert}, \citenamefont {Parks}, \citenamefont {Corkum},\
  and\ \citenamefont {Brabec}}]{bart_photonics20}%
  \BibitemOpen
  \bibfield  {author} {\bibinfo {author} {\bibfnamefont {G.}~\bibnamefont
  {Bart}}, \bibinfo {author} {\bibfnamefont {I.}~\bibnamefont {Schubert}},
  \bibinfo {author} {\bibfnamefont {A.}~\bibnamefont {Parks}}, \bibinfo
  {author} {\bibfnamefont {P.}~\bibnamefont {Corkum}}, \ and\ \bibinfo {author}
  {\bibfnamefont {T.}~\bibnamefont {Brabec}},\ }\href@noop {} {\bibfield
  {journal} {\bibinfo  {journal} {Journal of Physics: Photonics}\ }\textbf
  {\bibinfo {volume} {2}},\ \bibinfo {pages} {034007} (\bibinfo {year}
  {2020})}\BibitemShut {NoStop}%
\bibitem [{\citenamefont {Derouillat}\ \emph {et~al.}(2018)\citenamefont
  {Derouillat}, \citenamefont {Beck}, \citenamefont {P{\'e}rez}, \citenamefont
  {Vinci}, \citenamefont {Chiaramello}, \citenamefont {Grassi}, \citenamefont
  {Fl{\'e}}, \citenamefont {Bouchard}, \citenamefont {Plotnikov}, \citenamefont
  {Aunai} \emph {et~al.}}]{derouillat_cpc2018}%
  \BibitemOpen
  \bibfield  {author} {\bibinfo {author} {\bibfnamefont {J.}~\bibnamefont
  {Derouillat}}, \bibinfo {author} {\bibfnamefont {A.}~\bibnamefont {Beck}},
  \bibinfo {author} {\bibfnamefont {F.}~\bibnamefont {P{\'e}rez}}, \bibinfo
  {author} {\bibfnamefont {T.}~\bibnamefont {Vinci}}, \bibinfo {author}
  {\bibfnamefont {M.}~\bibnamefont {Chiaramello}}, \bibinfo {author}
  {\bibfnamefont {A.}~\bibnamefont {Grassi}}, \bibinfo {author} {\bibfnamefont
  {M.}~\bibnamefont {Fl{\'e}}}, \bibinfo {author} {\bibfnamefont
  {G.}~\bibnamefont {Bouchard}}, \bibinfo {author} {\bibfnamefont
  {I.}~\bibnamefont {Plotnikov}}, \bibinfo {author} {\bibfnamefont
  {N.}~\bibnamefont {Aunai}},  \emph {et~al.},\ }\href {\doibase
  10.1016/j.cpc.2017.09.024} {\bibfield  {journal} {\bibinfo  {journal}
  {Comput. Phys. Comm.}\ }\textbf {\bibinfo {volume} {222}},\ \bibinfo {pages}
  {351} (\bibinfo {year} {2018})}\BibitemShut {NoStop}%
\bibitem [{\citenamefont {Mironov}\ \emph {et~al.}(2025)\citenamefont
  {Mironov}, \citenamefont {Gelfer}, \citenamefont {Tupitsyn}, \citenamefont
  {Beck}, \citenamefont {Jirka}, \citenamefont {Klimo}, \citenamefont {Meuren},
  \citenamefont {Oberreit}, \citenamefont {Smorodnikova}, \citenamefont
  {Ta{\"\i}eb} \emph {et~al.}}]{mironov_pre25}%
  \BibitemOpen
  \bibfield  {author} {\bibinfo {author} {\bibfnamefont {A.}~\bibnamefont
  {Mironov}}, \bibinfo {author} {\bibfnamefont {E.}~\bibnamefont {Gelfer}},
  \bibinfo {author} {\bibfnamefont {I.}~\bibnamefont {Tupitsyn}}, \bibinfo
  {author} {\bibfnamefont {A.}~\bibnamefont {Beck}}, \bibinfo {author}
  {\bibfnamefont {M.}~\bibnamefont {Jirka}}, \bibinfo {author} {\bibfnamefont
  {O.}~\bibnamefont {Klimo}}, \bibinfo {author} {\bibfnamefont
  {S.}~\bibnamefont {Meuren}}, \bibinfo {author} {\bibfnamefont
  {G.}~\bibnamefont {Oberreit}}, \bibinfo {author} {\bibfnamefont
  {T.}~\bibnamefont {Smorodnikova}}, \bibinfo {author} {\bibfnamefont
  {R.}~\bibnamefont {Ta{\"\i}eb}},  \emph {et~al.},\ }\href@noop {} {\bibfield
  {journal} {\bibinfo  {journal} {Physical Review E}\ }\textbf {\bibinfo
  {volume} {112}},\ \bibinfo {pages} {055202} (\bibinfo {year}
  {2025})}\BibitemShut {NoStop}%
\bibitem [{\citenamefont {Radzig}\ and\ \citenamefont
  {Smirnov}(2012)}]{radzig}%
  \BibitemOpen
  \bibfield  {author} {\bibinfo {author} {\bibfnamefont {A.~A.}\ \bibnamefont
  {Radzig}}\ and\ \bibinfo {author} {\bibfnamefont {B.~M.}\ \bibnamefont
  {Smirnov}},\ }\href@noop {} {\emph {\bibinfo {title} {Reference data on
  atoms, molecules, and ions}}},\ Vol.~\bibinfo {volume} {31}\ (\bibinfo
  {publisher} {Springer Science \& Business Media},\ \bibinfo {year}
  {2012})\BibitemShut {NoStop}%
\bibitem [{\citenamefont {Popov}(2001)}]{popov-jetp01}%
  \BibitemOpen
  \bibfield  {author} {\bibinfo {author} {\bibfnamefont {V.}~\bibnamefont
  {Popov}},\ }\href@noop {} {\bibfield  {journal} {\bibinfo  {journal} {Journal
  of Experimental and Theoretical Physics}\ }\textbf {\bibinfo {volume} {93}},\
  \bibinfo {pages} {278} (\bibinfo {year} {2001})}\BibitemShut {NoStop}%
\bibitem [{\citenamefont {Tong}\ and\ \citenamefont {Lin}(2005)}]{tong_jpb05}%
  \BibitemOpen
  \bibfield  {author} {\bibinfo {author} {\bibfnamefont {X.}~\bibnamefont
  {Tong}}\ and\ \bibinfo {author} {\bibfnamefont {C.}~\bibnamefont {Lin}},\
  }\href@noop {} {\bibfield  {journal} {\bibinfo  {journal} {Journal of Physics
  B: Atomic, Molecular and Optical Physics}\ }\textbf {\bibinfo {volume}
  {38}},\ \bibinfo {pages} {2593} (\bibinfo {year} {2005})}\BibitemShut
  {NoStop}%
\bibitem [{\citenamefont {Bauer}\ and\ \citenamefont
  {Mulser}(1999)}]{bauer_pra1999}%
  \BibitemOpen
  \bibfield  {author} {\bibinfo {author} {\bibfnamefont {D.}~\bibnamefont
  {Bauer}}\ and\ \bibinfo {author} {\bibfnamefont {P.}~\bibnamefont {Mulser}},\
  }\href {\doibase 10.1103/PhysRevA.59.569} {\bibfield  {journal} {\bibinfo
  {journal} {Phys. Rev. A}\ }\textbf {\bibinfo {volume} {59}},\ \bibinfo
  {pages} {569} (\bibinfo {year} {1999})}\BibitemShut {NoStop}%
\bibitem [{\citenamefont {Golovanov}\ and\ \citenamefont
  {Kostyukov}(2020)}]{golovanov2020formula}%
  \BibitemOpen
  \bibfield  {author} {\bibinfo {author} {\bibfnamefont {A.~A.}\ \bibnamefont
  {Golovanov}}\ and\ \bibinfo {author} {\bibfnamefont {I.~Y.}\ \bibnamefont
  {Kostyukov}},\ }\href {\doibase https://doi.org/10.1070/QEL17309} {\bibfield
  {journal} {\bibinfo  {journal} {Quantum Electron.}\ }\textbf {\bibinfo
  {volume} {50}},\ \bibinfo {pages} {350} (\bibinfo {year} {2020})}\BibitemShut
  {NoStop}%
\end{thebibliography}%

\end{document}